\documentclass[fleqn,usenatbib]{mnras}

\usepackage{newtxtext,newtxmath}
\usepackage[T1]{fontenc}
\usepackage{ae,aecompl}
\usepackage{url}
\let\oldhref\href
\renewcommand{\href}[2]{\oldhref{#1}{\hbox{#2}}}
\usepackage{graphicx}	% Including figure files
\usepackage{amsmath}	% Advanced maths commands
\usepackage{amssymb}	% Extra maths symbols

\title[Post-starburst galaxies in the field]{The strong correlation between post-starburst fraction and environment}

\author[A. Paccagnella et al.]{A. Paccagnella,$^{1, 2}$\thanks{E-mail: angela.paccagnella@gmail.com (AP)}
B.Vulcani,$^{3,1}$\thanks{E-mail: benedetta.vulcani@inaf.it (BV)}
B.M. Poggianti,$^{1}$
A. Moretti,$^{1}$
J. Fritz,$^{4}$
\newauthor
M. Gullieuszik,$^{1}$
G. Fasano$^{1}$
\\
$^{1}$Department of Physics and Astronomy, University of Padova, Vicolo Osservatorio 3, 35122 Padova, Italy\\
$^{2}$INAF - Astronomical Observatory of Padova, 35122 Padova, Italy\\
$^{3}$School of Physics, University of Melbourne, VIC 3010, Australia \\
$^{4}$Instituto de Radioastronom\'\i a y Astrof\'\i sica, IRyA, UNAM, Campus Morelia, A.P. 3-72, C.P. 58089, Mexico\\
}

\date{Accepted XXX. Received YYY; in original form ZZZ}

\pubyear{2017}
\hypersetup{draft}
\begin{document}
\label{firstpage}
\pagerange{\pageref{firstpage}--\pageref{lastpage}}
\maketitle

\begin{abstract}
We examine a magnitude limited ($M_B\leq -18.7$) sample of post-starburst (PSB) galaxies at  $0.03<z<0.11$ in the different environments from the spectroscopic data set of the Padova Millennium 
Galaxy Group Catalog and compare their incidence and  properties  with those of passive (PAS) and emission line galaxies (EML).   PSB galaxies  have a quite precise life-time ($<1-1.5$Gyr), and they hold important clues for  understanding  galaxy evolution.  While the properties (stellar mass, absolute magnitude, color) of PSBs do not depend on environment,  their frequency increases going from single galaxies to binary systems to groups, both considering the incidence with respect to the global number of galaxies and to the number of currently+recently star-forming galaxies.
Including in our analysis the sample of cluster PSBs drawn from the WIde-field Nearby Galaxy-cluster Survey  presented in \citeauthor{Paccagnella2017}, we  extend the halo mass range covered and present a coherent picture of the effect of the environment on galaxy transformations.  
We find that the PSB/(PSB+EML) fraction steadily increases with halo mass going from 1\% in $10^{11} M_{\odot}$ haloes to $\sim 15$\% in the most massive haloes ($10^{15.5} M_{\odot}$).
{This provides evidence that processes specific to the densest environments, such as ram pressure stripping, are responsible for a large fraction of PSB galaxies in dense environments. These processes act on a larger fraction of galaxies than alternative processes leading to PSB galaxies in the sparsest environments, such as galaxy interactions.}%{ Our analysis does not allow us to definitely identify the main mechanisms responsible for the appearance of the  PSB spectrum. However, the higher incidence of PSB galaxies in the densest 
%environments suggests that the  processes operating in the most massive haloes, such as ram pressure stripping, are more effective in producing such population than the processes operating in the sparsest environments, most likely galaxy interactions.}
\end{abstract}


%%%%%%%%%%%%%%%%%%%%%%%%%%%%%%%%%%%%%%%%%%%%%%%%%%

%%%%%%%%%%%%%%%%% BODY OF PAPER %%%%%%%%%%%%%%%%%%

\section{Introduction}\label{sec:intro}
Post-starburst galaxies are  a distinct class of objects, whose main spectroscopic characteristics are strong Balmer absorption lines and no emission lines due to star formation. They were  identified for the first time by \citet{Dressler1983} and, since then, they have been largely analyzed also through extensive spectrophotometric modeling  \citep[e.g.][]{Couch1987,Newberry1990,Poggianti1996,Abraham1996, Poggianti1997,Bekki2001,Poggianti2004, Poggianti2004b}.  These galaxies have been found to have had a large burst of star formation (or had at least active star formation, in the weakest cases)  at some point during the $\sim$1 Gyr before the time of observation and then a sudden quench, and now they  show very little ongoing star formation \citep{Couch1987}. As a result, the stellar population in these galaxies is old enough that the massive, short-lived O- and B-stars have all died (showing no nebular emission lines), and young enough that  A-stars dominate the optical spectrum (showing strong Balmer absorption).  Given their properties, { different studies have suggested different origins for the post-starburst galaxies. It has been proposed that} these objects are in a transitional stage between star-forming blue cloud galaxies and quiescent red sequence galaxies \citep[e.g.][]{Caldwell1996, Zabludoff1996, Norton2001, Pracy2009,Zwaan2013, Yesuf2014, Pattarakijwanich2016, Paccagnella2017, Wu2014} and  therefore { might be} closely related to the ``green valley'' galaxies. Nonetheless, not all green valley galaxies have the spectroscopic signature of post-starburst galaxies \citep[e.g.][]{Wong2012,Vulcani2015}. These galaxies may { also} represent a key stage in the formation of a large fraction of the passive population \citep{Wild2016},
{ be rejuvenating from the quiescent population \citep[e.g.][]{Dressler2013, Abramson2013} or returning to the star forming population \citep{Rowlands2015, French2015, Wild2016}. 
}

Two main scenarios have been proposed to explain the formation of post-starburst galaxies, which most likely describe their existence  in different environments. 
The first scenario justifies the presence of post-starbursts  in clusters, where \cite{Dressler1982} first observed them, hypothesizing that they lay exclusively in over-dense regions.  In this picture,  the main mechanism is a cluster-related process such as ram-pressure stripping. Gas-rich star-forming galaxies fall into  clusters and their gas is removed by the interaction with the hot intra-cluster medium, suddenly quenching their star formation \citep{Dressler1982, Couch1987, Dressler1992, Poggianti1999, Balogh2000, Tran2003,Tran2004,Tran2007, Poggianti2009, Paccagnella2017}. The resulting post-starburst galaxies would still resemble disk galaxies, since nothing apart from gas loss would disturb their morphologies.  This explanation holds also for post-starburst galaxies in massive groups, where ram pressure is still effective \citep[e.g.][]{Poggianti2009}. 
{Recently, \cite{Socolovsky2018} found that post-starburst  galaxies in clusters are preferentially produced by a rapid truncation following an extended period of star formation or after a minor starburst, rather than gas depletion after a major starburst. They conclude that this implies that environmental mechanisms may quench galaxies without triggering any significant burst of star formation.
} 

The second scenario  was introduced to explain the presence of post-starburst galaxies  in lower density regions such as poor groups and the field \citep{Blake2004, Goto2005, Zabludoff1996, Quintero2004}. These  galaxies might be associated with galaxy interactions, mergers or AGN feedback \citep{Bekki2001, Quintero2004, Blake2004,Goto2005, Hogg2006, Coil2011, Mahajan2013, Trouille2013, Yesuf2014}. In this picture, when gas-rich galaxies go through a merger phase, the gas is disturbed and collapses to form stars, leading to a large scale starburst. The same disturbance also channels the gas to the galactic center, resulting in both a central starburst and AGN activity. Then, either the  galaxies run out of gas and stop forming stars, or the rest of the gas receives enough heating from  supernova or AGN feedback that it becomes too hot to collapse further or is expelled altogether. Either way, these galaxies go through a starburst phase that stops quickly. Within this picture, their morphologies should be disturbed and show prominent signs of recent interactions such as tidal tails \citep{Zabludoff1996, Yang2004, Yang2008, Blake2004, Tran2004, Goto2005, Pracy2009}, { even though the faint merger remnant features might decay much more rapidly than the post-starburst features \citep{Pawlik2016}.} 

Spatially resolved kinematics and stellar population analysis  suggest that mergers are the dominant mechanism in the field \citep{Pracy2012, Pracy2013}. In particular,  Balmer {absorption} line enhancement in the central regions of post-starburst galaxies might be indicative of young central merger-induced star bursts \citep{Swinbank2011, Whitaker2012, Pracy2013}.  

%\bbv{non sono sicura che quello che segue centri..}
Another type of galaxies with spectral characteristics similar to post-starbursts, but with important differences, are those with both emission-lines and strong $\rm H\delta$ (>4\AA) in absorption, the so called ``e(a)'' galaxies \citep{Dressler1999, Poggianti1999,
Poggianti2000,Miller2001, Yesuf2014}. \cite{Poggianti1999} suggested that this kind of
galaxies are in fact dust-enshrouded starburst galaxies, and \cite{Poggianti2000} showed that this type of spectra are in fact common among Luminous and Ultraluminous Infrared galaxies. In this picture, the ongoing star formation is hidden behind a large amount of dust and gas and therefore the emission lines are strongly extincted. Balmer absorption lines, on the other hand, originate from A-stars, which are long-lived enough to migrate out of the star- forming gas cloud and become visible. % \bbv{***}
It is instead still debated what fraction of the post-starburst spectra (with no emission lines) are
dust-enshrouded starbursts \citep{Goto2004, Chang2001, Dressler2009, Nielsen2012}.
A number of galaxies with both Balmer strong absorption and emission lines have also been found to be ``shocked Poststarburst galaxies'', in which the nebular lines are excited via shocks, that have been interpreted as an earlier phase in galaxy transformations than traditional post-starburst galaxies \citep{Alatalo2016, Alatalo2017, Baron2018}.

Few studies have started comparing  samples of post-starbursts in clusters and in the field.
{ In the local Universe, \cite{Zabludoff1996, Hogg2006} found that post-starburst galaxies are found in the same environments as star- forming galaxies \citep[see also][]{Quintero2004, Blake2004, Goto2005}. In clusters, \cite{Hogg2006} also found a marginal larger ratio of post-starburst to star-forming galaxies within the virial radius than outside, while \cite{vonderLinden2010} did not detect any variation with radius. }

 \cite{Paccagnella2017} {  presented the first complete characterization of post-starburst galaxies in clusters at low-redshift. } They investigated the incidence, properties of the stellar populations  and the spatial location of post-starburst galaxies within clusters, in a sample at $0.04<z<0.07$, based on the WIde-field Nearby Galaxy-cluster Survey (WINGS) \citep{Fasano2006, Moretti2014} and OmegaWINGS \citep{Gullieusik2015, Moretti2017} data. They found that { considering a magnitude limited sample including all galaxies with an apparent V magnitude brighter than 20}, post-starbursts  are $7.2\pm0.2\%$ of  cluster members within 1.2 virial radii. Their frequency raises from the outskirts toward the  cluster cores and from the least toward the most luminous and massive clusters.  
{ Post starburst magnitudes, stellar masses, morphologies and  colors typically show an intermediate behavior between passive and emission line galaxies, indication of a population in transition from being star forming to passive. 
These results, together with the phase-space analysis and velocity dispersion profile, made \citeauthor{Paccagnella2017}  conclude that post starburst galaxies represent 
cluster galaxies that accreted onto  clusters quite  recently. 
Ram-pressure stripping was proposed as the main responsible for the rapid truncation of star formation required to explain the spectral features observed in post starburst galaxies. } %\bbv{vedere se lasciarlo qui}
Overall, in the local universe, a  coherent picture of the properties and incidence of post-starbursts in the different environments in an homogeneous sample is still lacking.

{A larger number of studies focused on post-starburst galaxies at intermediate redshift, reporting contradictory results. 
\cite{Balogh1997, Balogh1999} found no enhancement of post-starburst galaxies in  clusters at z $\sim$0.3. In contrast,  \cite{Dressler1999, Poggianti1999} did find a clear increase in the incidence of post-starbursts at z $\sim$0.5. Subsequently, other works showed evidence for both cluster-triggered post-starbursts \citep{Tran2004, Poggianti2009}, and  suppression of post-starburst in groups \citep{Yan2009}. 
Focusing on one cluster at $z\sim0.55$, \cite{Ma2008} found that  E+A galaxies are found exclusively within the ram-pressure radius of the cluster and that their  colors confirm that they represent the transition phase between emission-line and absorption- line galaxies.
\citet{Poggianti2009} found that the incidence of $k+a$ galaxies depends strongly on environment: the poststarburst 
%reside preferentially 
fraction is higher
in clusters and in a subset of  groups  showing a low fraction of [OII] emitters.  In contrast, $k+a$ galaxies are 
proportionally less frequent
%numerous 
in the field, in poor groups and groups with a high [OII] fraction. 
Also \citet{Dressler2013}, found that post starburst galaxies strongly favor denser environments and that their incidence increases going from the field to groups and clusters.
\cite{Vergani2010} found that at $z=0.48-1.2$  post-starburst galaxies are morphologically a heterogeneous population with a similar incidence of bulge-dominated and disky galaxies. Their spatial location overlap that of quiescent galaxies on a physical scale of $\sim$2-8 Mpc.
\cite{Wu2014} showed how the prevalence of post-starburst galaxies in clusters correlates with the dynamical state of the host cluster, as both galaxy mergers and the dense ICM produce post-starburst galaxies.

}

%In the local universe, 
%The  studies performed so far investigated either samples of post-starbursts in clusters or in the field,  without building a coherent picture and quantifying  the incidence of  this population in different   environments in the local universe  in an homogeneous sample.

In this work we extend the analysis presented in \cite{Paccagnella2017}, by characterizing the incidence and stellar population properties of post-starburst galaxies in lower-density environments, specifically in groups, binaries and single systems, and contrast their incidence to the same population in  clusters. 
We base our analysis on the Padova-Millennium Galaxy and Group Catalogue \citep[PM2GC,][]{Calvi2011},  a spectroscopically complete sample of galaxies at $0.03 \leq z \leq 0.11$ with   $M_B \leq -18.7$. 
This  group catalog has been already successfully exploited to investigate the role of the group environment in driving galaxy transformations \citep[e.g.][]{Vulcani2011, Calvi2012, Calvi2013,Poggianti2013, Guglielmo2015, Vulcani2015}.

\cite{Vulcani2015} already briefly investigated the incidence of post-starburst galaxies  in a mass limited sample ($\log(M_\ast/M_\odot)\geq 10.25$) of galaxies drawn from the PM2GC. They found that, in the general field,
4.5\% of all galaxies   can be classified as post-starbursts. However, when considering the relative number of post-starburst and blue galaxies (used as a proxy of currently star-forming galaxies), this fraction is much lower in binaries and singles (13 $\pm$ 5\%) than
 in groups (22 $\pm$ 6\%), suggesting a higher efficiency of sudden quenching in groups compared to lower mass haloes.  Here we build on their findings, { characterizing in detail the stellar population properties of the post-starburst galaxies, and exploiting a more accurate definition of environment. We also combine the PM2GC and WINGS+OmegaWINGS samples, with the aim of } characterizing the incidence of post-starbursts over a wide range of halo masses, namely  $10^{11} M_\odot < M_{halo}< 10^{15.5} M_\odot$.

Throughout the paper, we adopt a \cite{Salpeter1955} initial mass function (IMF) in the mass range 0.1-125 M$_{\odot}$. The cosmological constants assumed are $\Omega_m=0.3$, $\Omega_{\Lambda}=0.7$ and H$_0=70$ km s$^{-1}$ Mpc$^{-1}$.

\section{The data samples}\label{sec:data}

\subsection{The field sample}
{ The main sample of this analysis is }
the PM2GC \citep{Calvi2011}, a spectroscopically complete sample of galaxies at $0.03 \leq z \leq 0.11$ brighter than $M_B = -18.7$. 
The galaxy sample was built on the basis of the Millennium Galaxy Catalogue (MGC) \citep{Liske2003, Cross2004,Driver2005}, a deep and wide B-imaging survey (B<20) along an equatorial strip of $0.6 deg \times 73 deg$ covering an area of $\sim38 deg^2$ obtained with the Isaac Newton Telescope (INT) and consisting of 144 overlapping fields.

A sample of galaxies included in the MGCz catalog, the spectroscopic extension of the MGC, was selected. $\sim$86\% of the MGCz spectra of the sample come from the SDSS \citep[$\sim$2.5 \AA{} resolution, { median signal-to-noise per pixel 15,} ][]{Abazajian2003}, 12\% from the 2dF Galaxy Redshift Survey \citep[2dFGRS, $\sim$ 9 \AA{} resolution, { median signal-to-noise per pixel 13,}][]{Colless2001} and 2\% from the 2dF follow-up obtained by the MGC team \citep[][]{Driver2005}. The fibre diameters are 3$^{\prime\prime}$ for the SDSS and 2.16$^{\prime\prime}$ for the 2dF setup, corresponding to the inner 1.3 to 6 kpc of the galaxies.  While SDSS spectra were already flux-calibrated, for all 2dF spectra a relative flux calibration was performed. We  used the response curve provided by the 2dF team. We refined the curve using the ratio between the 2dF spectra and the corresponding SDSS spectra for objects with both spectra available, as done  by \citet{Cava2009}. 

The approach adopted to identify galaxy groups is based on a  friends-of-friends (FoF) algorithm \citep{Calvi2011}.
 A galaxy is considered as a  member of a group if its spectroscopic redshift lies within $\pm 3\sigma$ (velocity dispersion) from the median group redshift and if it is located within a projected distance of 1.5R$_{200}$ from the group geometrical center.  R$_{200}$ is defined as the radius delimiting a sphere with interior mean density 200 times the critical density of the universe at that redshift, and is commonly used as an approximation of the group virial radius.
The R$_{200}$ values are computed from the velocity dispersions as in \citet{Poggianti2006}.
The result is a catalog including 176 groups of galaxies with at least three members, in the redshift range $0.04\leq z\leq 0.1$. The catalog contains 1057 galaxies that represent 43\% of the entire general field population in the same redshift range. 
The vast majority of the groups contain fewer than 10 members, 63\% have less than 5 members and 43\% have only three members. The median redshift  of the sample is $z=0.0823$. The range of velocity dispersion is between 100 km/s and 800 km/s, with a median  value of $\sigma=191.8$ km/s

Galaxies  not satisfying the group membership criteria at $0.03\leq z\leq0.11$ and with no friends (1141) were classified as field-single, those with only one friend within 1500 km/s and 0.5 $h^{-1} Mpc$ (490) were classified as field-binary.
All galaxies in the environments described above are
collected in the ``general field'' sample.

\subsection{Probing the environments with mock catalogs in the PM2GC} \label{sec:sims} 
Due to projection effects, in observations any method adopted to identify groups and their members produces some fraction of false positives, that is  interlopers selected as members that do not belong to a group, and false negatives, that is true members missed by the selection procedure. 

To test the performance of the FoF algorithm adopted to define the different environments \citep{Calvi2011},  we exploit mock catalogs drawn from the Millennium Simulation \citep{Springel2005}. We also use the same simualtions to assign estimates of halo masses to the PM2GC structures, as described in the following.

\subsubsection{The simulation}
We base our analysis on the publicly available  catalogs
from the semi-analytic model of \citet{DeLucia2007} run on the Millennium Simulation \citep{Springel2005}. The simulation  has a spatial resolution of 5 $h^{-1}$ kpc, uses $10^{10}$ particles of mass $8.6 \times 10^{8} h^{-1} \, M_\odot$ and traces the evolution of the matter distribution in a cubic region of the universe of 500 $h^{-1}\, Mpc$, on a side from $z = 127$ until $z = 0$.

%We take advantage of publicly available galaxy catalogs
%from semi-analytic models run on the Millennium Simulation \citep{Springel2005}. This uses $10^{10}$ particles of mass $8.6 \times 10^{8} h^{-1} \, M_\odot$ to trace the evolution of the matter distribution in a cubic region of the universe of 500 $h^{-1}\, Mpc$, on a side from $z = 127$ until $z = 0$, and has a spatial resolution of 5 $h^{-1}$ kpc.

The \citet{DeLucia2007} model builds on the methodology and prescriptions originally introduced by \citet{Springel2001,DeLucia2004a} and \citet{Croton2006}. 
We refer to the original papers for  details. 
%In this work we use the semi-analytic model discussed in \citet{DeLucia2007}, which builds on the methodology and prescriptions introduced by \citet{Springel2001,DeLucia2004a} and \citet{Croton2006} and has been the first variant of the ``Munich'' models family that has been made publicly available. It includes prescriptions for supernova-driven winds, follow the growth of supermassive black holes, and include a phenomenological description of AGN feedback. The model is calibrated to reproduce the $z=0$ luminosity function in the K- and J -bands. We refer to the original paper for more details.
We exploit the extractions presented in \citet{Vulcani2014}, which were used to reproduce the PM2GC  field. 10 portions of ``simulated sky'' corresponding to square boxes of $\sim 38 deg^2$ (30$\times$30 Mpc) at $z=0.06$ each were selected. The boxes extracted were 323 physical Mpc deep. 
No pre-selection on halo mass was applied to these boxes.

%We exploit the extractions presented in \citet{Vulcani2014}, which were used to reproduce the observed field of the PM2GC selecting portions of ``simulated sky'' corresponding to square boxes of $\sim 38 deg^2$ (30$\times$30Mpc) at $z=0.06$. The boxes extracted were 323 physical Mpc deep. 
%To account for cosmic variance, 10 simulated field samples were selected and will be used separately. No pre-selection on halo mass was applied to these boxes.

The model provides information on  halo masses,  virial radii, velocity dispersions, number of galaxies, galaxy  stellar masses, galaxy positions in the box,  and galaxy magnitudes.

In what follows, we refer to an output value of the simulations either as a ``simulated'' or a ``sim-projected'' quantity. The former are the 3D estimates provided by the simulation; the latter are computed from the simulation with the same methods that has been used observationally and are projected on the $xy$ plane.

%In the following, we refer to an output value of the simulations either as a ``sim-projected'' or a ``simulated'' quantity. Sim-projected quantities (number of galaxies, velocity dispersion, etc.) are computed from the simulation with the same methods that has been used observationally and are projected on the $xy$ plane, while simulated quantities are the 3D estimates provided by the simulation.

%Velocity dispersions have been computed using all galaxies within R$_{200}$ and more massive than $M_\ast \sim 5 \times 10^{8}M_{\odot}$, that is the resolution limit of the simulation.

%The model provides information on the halo mass, the virial radius, the stellar masses, the $xyz$ coordinates in the box and the velocity dispersions.
%When using the sim-projected sample,  galaxies are projected on one plane.
%Velocity dispersions have been computed using all galaxies within R$_{200}$ and more massive than $M_\ast \sim 5 \times 10^{8}M_{\odot}$, that is the resolution limit of the simulation.

To reduce the biases between observations and simulations, 
we compute stellar masses in an homogeneous way for the sim-projected and the observed sample. We therefore 
adopt the \citet{BelldeJong2001} prescription and correlate the stellar mass-to-light ratio with the optical colors of the integrated stellar population:
\begin{equation}
\log(M/L_B) = -0.51+1.45(B-V) 
\end{equation}
valid for a Bruzual \& Charlot model with solar metallicity
and a \citet{Salpeter1955} IMF (0.1-125 M$_{\odot}$).  The typical uncertainty on mass estimates is 0.2-0.3 dex. 
%The rest-frame Vega magnitudes include the effect of  dust in the \citet{Buser1978} system, calculated using the models of \citet{Bruzual2003}. . 
For the sim-projections, we  use the rest-frame Vega magnitudes from the models;\footnote{These magnitudes include the effect of  dust in the \citet{Buser1978} system, calculated using the models of \citet{Bruzual2003}.} for the PM2GC, we use the B-band photometry taken from the MGC, and the rest-frame B-V color computed from the Sloan g-r color corrected for Galactic extinction \citep[see][for details]{Calvi2011}.
\citet{Vulcani2014} performed a comparison between the    masses provided by the model and those obtained following the \citet{BelldeJong2001} formulation, highlighting a good agreement, with an absolute median difference of 0.08 dex.  %Similar values are found also for the PM2GC.

%The model also provides rest-frame Vega magnitudes, which include the effect of  dust in the \citet{Buser1978} system, calculated using the models of \citet{Bruzual2003}, and stellar masses. Nonetheless, in order to reduce the biases between observations and simulations, we  compute stellar masses in an homogeneous way for the sim-projected and the observed sample. We 
%follow the \citet{BelldeJong2001} relation, which correlates the stellar mass-to-light ratio with the optical colors of the integrated stellar population:
%\begin{equation}
%\log(M/L_B) = -0.51+1.45(B-V) 
%\end{equation}
%valid for a Bruzual \& Charlot model with solar metallicity
%and a \citet{Salpeter1955} IMF (0.1-125 M$_{\odot}$).  The typical uncertainty on mass estimates is 0.2-0.3 dex. 
%For the sim-projections, we  use the magnitudes from the models; for the PM2GC, we use the B-band photometry taken from the MGC, and the rest-frame B-V color computed from the Sloan g-r color corrected for Galactic extinction.
%For the sim-projections, as a consistency test, \citet{Vulcani2014} performed a comparison between these stellar masses and the ones provided by the code. They found a good agreement, with an absolute median difference of 0.08 dex.  Similar values are found also for the PM2GC.

\subsubsection{Contamination and incompleteness of the sim-projected groups}\label{sec:contamination}
To estimate the contamination and incompleteness in our observed group catalog, we run our FoF algorithm on each of the ten extractions of the sim-projected sample, limited at M$_B$=-18.7, separately, and identify sim-projected-groups, sim-projected-binary systems and sim-projected-single galaxies.

We then compare these outputs to the catalogs of structures obtained from the simulated samples using the real memberships.
The  number of systems in the two different samples, averaged over the 10 fields, are given in Tab. \ref{tab:modcat}. It appears evident that the FoF recovers many more groups than the real ones, suggesting a high level of contamination in the observed catalog. 
\begin{table}
\begin{center}
\caption{Mean number of groups, binary systems and single galaxies in the simulated and sim-projected samples. \label{tab:modcat}}
\begin{tabular}{lrrr}
  \multicolumn{1}{c}{Catalog} &
  \multicolumn{1}{c}{Groups} &
  \multicolumn{1}{c}{Binaries} &
  \multicolumn{1}{c}{Singles} \\
\hline
  simulated & 87$\pm$1 & 135.5$\pm$0.5 & 1893.3$\pm$0.5\\
  sim-projected & 165.7$\pm$0.7 & 212.5$\pm$0.4 & 969.2$\pm$0.6\\
\hline
\end{tabular}
\end{center}
\end{table}

We can call $N_{\rm sim}$ 
the number of real members given by the simulated sample,  $N_{\rm sp}$ 
the number of members recovered by the FoF, $N_{\rm true}$ the intersection between $N_{\rm sim}$ and $N_{\rm sp}$ and $N_{\rm i}=N_{\rm sp}-N_{\rm true}$ the number of interlopers. 
The ``contamination'' is defined as the fraction of interlopers: $f_{\rm i}=N_{\rm i}/N_{\rm sp}$. 
In the sim-projected samples, more than 50\% of the sim-projected groups have a contamination higher than 0.5.

\begin{figure}
\centering
 \includegraphics[scale=0.42]{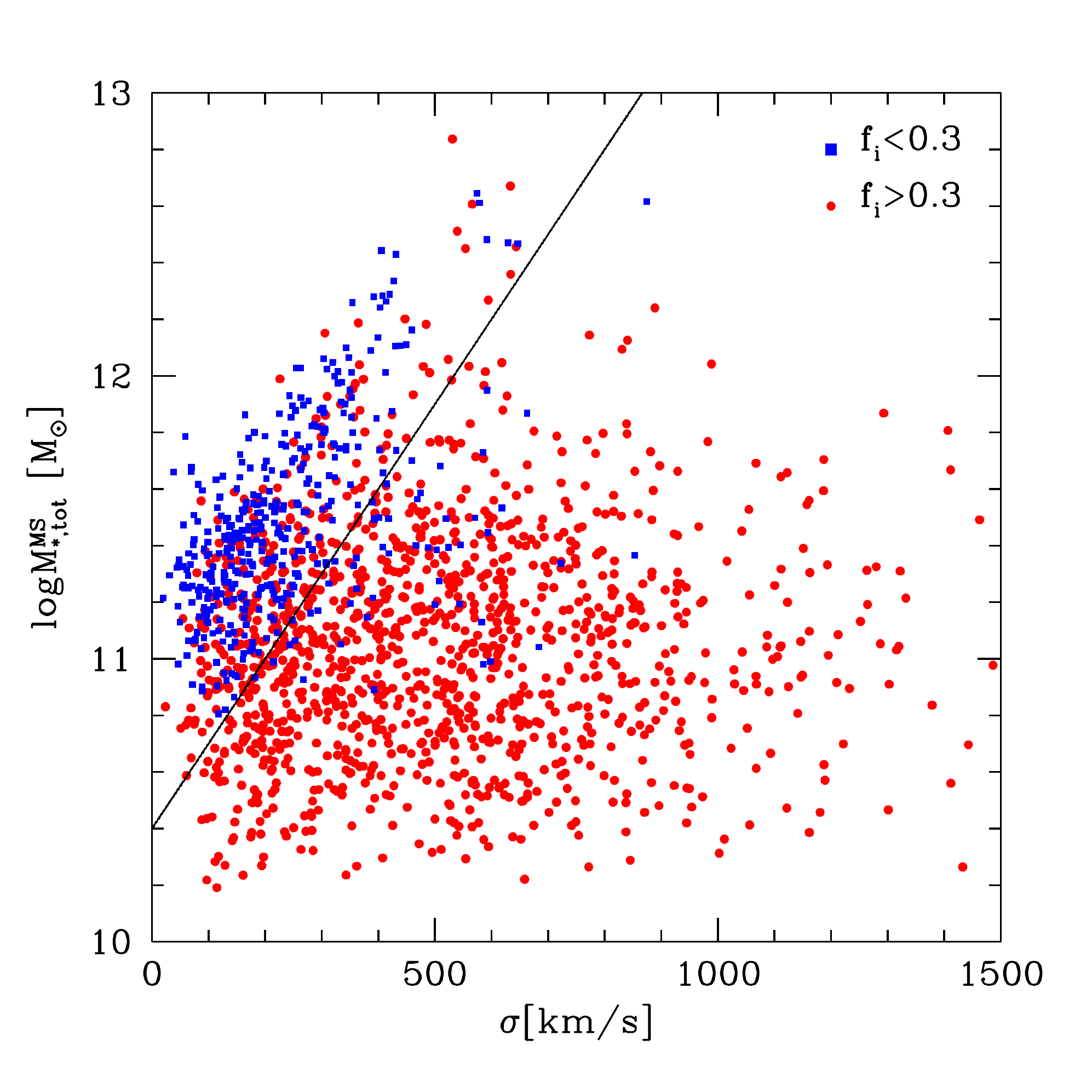}
 \vspace{-20pt}
 \caption{$\log (M_{\ast, tot}^{MS}/M_\odot)-\sigma$ relation for the ten samples of sim-projected-groups. Blue and red dots indicate  groups with a fraction of interlopers smaller and bigger than 0.3, respectively. The black line traces the locus where $\log(M_{*,tot}^{MS}/M_\odot=0.003\cdot\sigma+10.40$ (Eq.\ref{eq:msig}).  \label{sig-mass}}
\end{figure}

Figure \ref{sig-mass} shows 
the velocity dispersion of sim-projected groups $\sigma$ as a function of the total stellar mass of group members, $\log (M_{*,tot}^{MS}/M_\odot)$. The latter has been computed including only galaxies brighter than the PM2GC magnitude limit of $M_B=-18.7$.  While groups with $f_{\rm i}\geq 0.3$ are spread all over the plane, a quite tight correlation emerges for groups with $f_{\rm i}<0.3$,which are preferentially found at 
\begin{equation}
\label{eq:msig}
\log(M_{*,tot}^{MS}/M_\odot)>0.003\cdot\sigma+10.40
\end{equation}
Groups above this line have a fraction of interlopers of $\sim 0.2$, and from now on we will refer to them as ``real'' groups. In contrast,  groups below this line (from now on ``fake'' groups) have a fraction of interlopers of $\sim 0.6$.

The ``incompleteness'' is defined as the number of lost members $N_{\rm lost}=N_{\rm true}-N_{\rm sp}$. In the sim-projected samples, 7\% of the sim-projected groups have an incompleteness higher than 0.5.
The incompleteness will be used in the following Section to determine the halo mass of sim-projected groups.

\subsubsection{Halo mass estimates for the sim-projected groups}
\label{sec:halomass}

We assign halo masses to the sim-projected-groups with at least three members. 
For each sim-projected group, we select the simulated halo 
 that includes the majority of its members. 
 Then, we calculate the incompleteness of the system. 
If $N_{\rm lost}>N_{\rm sp}$, we select the second most common simulated halo present in the sim-projected group and recalculate the incompleteness. 
If $N'_{\rm lost}<N_{\rm lost}$, with $N'$ indicating the second simulated halo, we reject the first halo and assume that the second one better corresponds to the sim-projected group and iterate the process, otherwise we keep the first one. At the end, we assign to each sim-projected group the mass of the simulated halo that better represents it. 
We note that in 38\% of the sim-projected groups, all galaxies belong to different simulated halos. In these cases we assign to the group the halo mass of the halo where the most massive galaxy (MMG) is located. 

For galaxies in the sim-projected binary systems, if the two members do not belong to the same simulated halo, we use the mass of the halo hosting the MMG.
For single galaxies the associated halo mass is the one of the simulated halo that contains the galaxy. 
In what follows, we will refer to this halo mass estimate as $M_{halo}^{MS}$. 

\subsubsection{The Total Stellar-to-halo Mass Relation}
\begin{figure}
\vspace{-10pt}
\centering
 \includegraphics[scale=0.55]{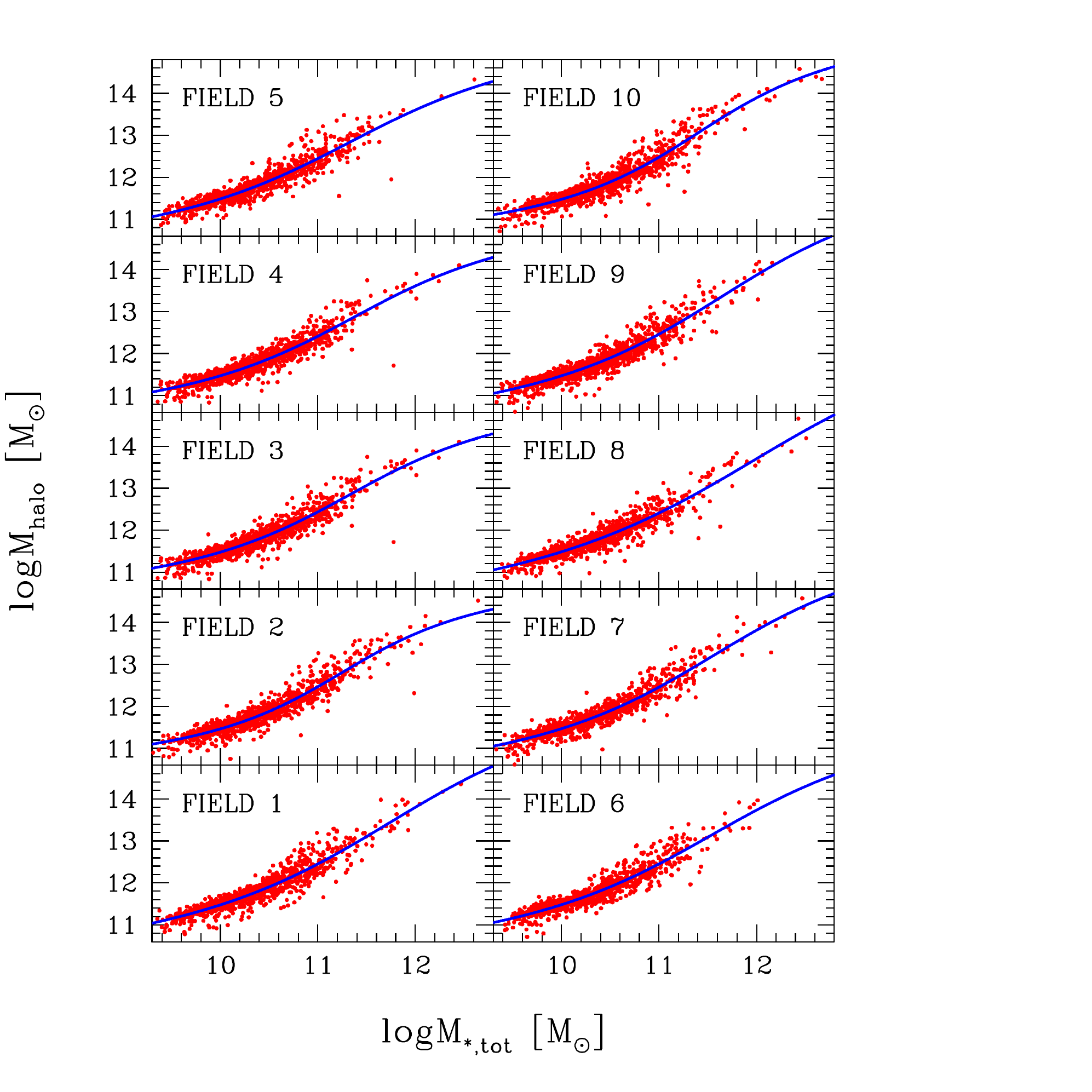}
 \caption{Main halo mass as a function of total stellar mass  for groups in the 10 sim-projected fields (red). Blue lines are the best fit results obtained using equation \ref{eq2}. \label{fit2}}
\vspace{-10pt}
 \end{figure}

Our aim is to  derive an halo mass estimate for each system in our observed group catalog. Therefore, we have to  link  our observables to  a quantity derived from simulations. 
As demonstrated in \cite{Yang2003,Yang2005a,Yang2007,Yang2008}, the mass of a dark matter halo associated with a group is tightly correlated with the total stellar mass of member galaxies. Such relation holds also in our sim-projected samples, as illustrated in Figure \ref{fit2}, where the correlations between the halo mass $\log(M_{halo}^{MS})$ and the total stellar mass $\log M_{\ast, tot}^{MS}/M_\odot$ for the 10 sim-projected fields are shown.
Real groups,\footnote{Including also fake groups would increase the scatter in the determination of the halo masses.} binaries and single systems are here considered all together. 
We fit the relation in each sim-projected sample using an analytic equation with four free parameters
\begin{equation}
\log M_{halo}^{tot}=a+d\left[arctan\left(\frac{\log M_{*,tot}^{MS}-b}{c}\right) \right]
\label{eq2}
\end{equation}
implementing a nonlinear least-squares Marquardt-Levenberg algorithm \citep{Marquardt1963}. 
The typical scatter of the relation  increases with halo mass, but is always $<0.4$ dex along the whole mass range.
Clearly, the scatter will result in errors in the inferred group masses. 
We then average the outputs of the ten fields, obtaining the following values: a=13.13$\pm$0.05, b=11.51$\pm$0.04, c=1.59$\pm$0.04, d=2.19$\pm$0.07.

\subsubsection{Halo masses for the PM2GC}
The procedure described above has been applied to groups identified by the same group finder algorithm run on the observations, thus already contains the scatter introduced with projection effects and interlopers on the estimation of the final halo masses. We can now apply Eq.\ref{eq2} to the observed sample, in order to  obtain  halo mass estimates for the PM2GC structures.

Before doing so,  we  have to consider that the observed sample is affected by survey geometry. Indeed, a group whose projected area straddles one or more survey edges may have members that fall outside of the survey area, thus causing an incompleteness that affects the mass estimate of the group.
The geometry used to quantify this issue is a simplification of the real one. Following \citet{Yang2007},  we  correct using a rectangular mask that contains the whole field of view.
To start, we estimate the total stellar mass\footnote{To take  into account the different cosmology adopted in the PM2GC and in the Millennium Simulation, we first apply a correction of 0.09 to the PM2GC magnitudes, which translates into a factor of $\sim0.032$ on stellar masses.} for each group without taking into account edge effects. We then randomly distribute 200 points within the corresponding halo radius R$_{200}$.
Next we apply the rectangular mask and compute the number of points that fall within the field region, N$_{r}$, and define f$_{edge}$=N$_{r}$/200 as a measure for the volume of the group that lies within the survey edges. Afterwards, we correct the total stellar mass by 1/f$_{edge}$ to take into account the ``missing members'' outside the edges.
This correction  works well for groups with f$_{edge} > 0.6$, that in the PM2GC sample constitute the 96\% of the whole sample.

Finally, we calculate the dark matter halo masses for the whole PM2GC catalog (groups, binaries and single galaxies) exploiting equation \ref{eq2} and flag as real those groups with $\log M_*>0.003\cdot \sigma+10.4$ that, according to the statistics derived in section \ref{sec:halomass}, should be the ones with the lower fraction of interlopers.

\begin{figure}
 \centering
 \includegraphics[scale=0.35]{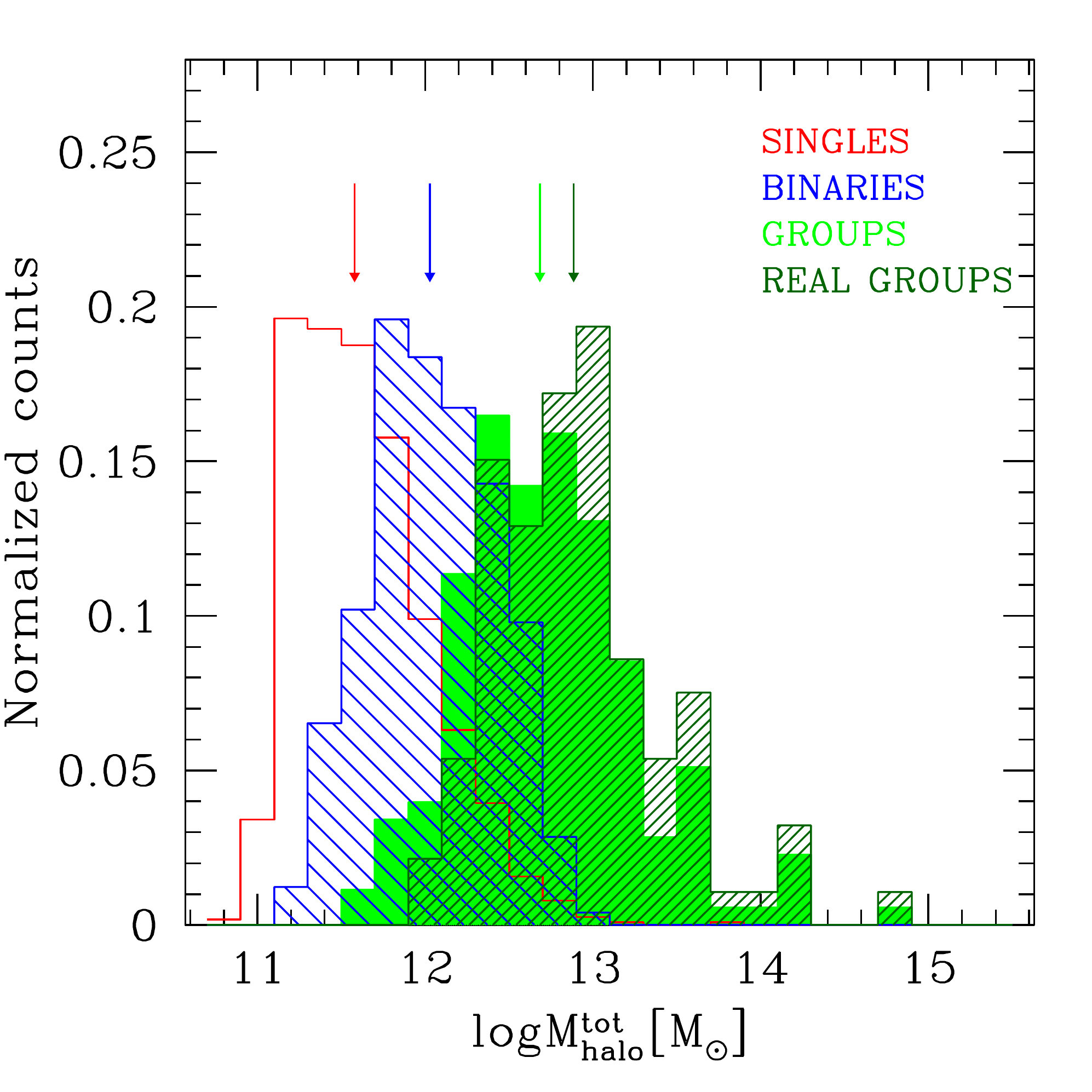}
 \caption{Normalized halo mass distribution for the PM2GC structures, obtained from eq.\ref{eq2}. Single galaxies  are shown in blue (hatched histogram), binary systems in red (empty histogram) and groups in green (filled histogram and hatched histogram for real groups). \label{fig:mdpm}}
\end{figure}

Figure \ref{fig:mdpm}  shows the distribution of the inferred halo masses for single galaxies, binary systems and groups in the PM2GC. Single galaxies span a halo mass range between $10^{10.9}M_\odot$ and $10^{13.9} M_\odot$, with a median value of $10^{11.58\pm 0.02} M_\odot$. Binary systems  span a halo mass range between $10^{11.2} M_\odot$ and $10^{12.9} M_\odot$, with a median value of $10^{12.03\pm 0.03} M_\odot$. Groups span a halo mass range between $10^{11.6} M_\odot$ and $10^{14.8} M_\odot$, with a median value of $10^{12.68\pm 0.05} M_\odot$. If we consider only real groups, the halo mass ranged spanned by the systems narrows, being $10^{12.0}-10^{14.8} M_\odot$, with a median value of $10^{12.88\pm 0.04} M_\odot$.

%In Appendix \ref{sec:sims} we test the performances of our FoF algorithm, exploiting mock catalogs drawn from the Millennium Simulation \citep{Springel2005} and identify the rate of incompleteness and contamination in our groups. We also develop  a functional formulation to compute halo mass estimates to the structures identified in the  PM2GC from observed quantities. 
The entire catalog of the halo mass estimates is described in Appendix \ref{sec:cat} and is given in the online version of the article.

%Our groups/binaries/singles span a halo mass range of $10^{11.6}-10^{14.8} M_\odot$ / $10^{11.2}-10^{12.9} M_\odot$ / $10^{10.9}-10^{13.9} M_\odot$, with a median value of $10^{12.68\pm 0.05} M_\odot$ / $10^{12.03\pm 0.03} M_\odot$ / $10^{11.58\pm 0.02} M_\odot$, respectively.
%
%When we consider only "real" groups, that is groups with a contamination from interlopers $f_{\rm i}<0.3$ (see Appendix \ref{sec:sims}), the halo mass range narrows ($10^{12.0}-10^{14.8} M_\odot$ with median value of $10^{12.88\pm 0.04} M_\odot$). This corroborates the hypothesis that the groups labeled as fake ($f_{\rm i}\geq 0.3$) are actually smaller systems contaminating the group sample. 

\subsection{The cluster sample}
{ The secondary sample of our analysis is}
 the WIde-field Nearby Galaxy-cluster Survey (WINGS) \citep{Fasano2006, Moretti2014} and its recent extension, OmegaWINGS \citep{Gullieusik2015, Moretti2017}, which quadruples the area covered. The two surveys together observed 76 clusters X-ray selected from ROSAT All Sky Survey data \citep{Ebeling1996,Ebeling1998,Ebeling2000} at $0.04<z<0.07$. The clusters span a wide range in X-ray luminosity ($L_X\sim 0.2-5 \times 10^{44}$ erg/s) and velocity dispersion ($\sigma$ typically between 500 and 1100 km s$^{-1}$). They therefore probe the massive tail of the halo mass distribution. 
{ Details on the target selection, redshift determinations, geometrical and magnitude incompleteness can be found in  \cite{Paccagnella2017}. }
{ Halo masses have been estimated from the cluster velocity dispersion $\sigma_{cl}$, by means of the virial theorem, following \citet{Poggianti2006}
\begin{equation}
M_{200}=1.2\cdot 10^{15}\left(\frac{\sigma_{cl}}{1000  km/s}\right)^3 \frac{1}{\sqrt{(\Omega_{\Lambda}+\Omega_M(1+z_{cl})^3)}}h^{-1} \quad 
(M_{\odot})
 \label{eq:m200}
\end{equation}
}

%Exploiting these data,  \cite{Paccagnella2017} presented the first complete characterization of PSB galaxies in clusters at low-redshift { and we defer the interested reader to that paper.} 
%Galaxy properties were derived by means of the same spectrophotometric code described in Sec. \ref{sec:gal_prop} and the same classification was adopted to identify the different spectral types (Sec. \ref{sec:spec_class}).
%As mentioned in Sec. \ref{sec:intro}, considering an apparent magnitude-limited sample ($V<20$), \citeauthor{Paccagnella2017} found that PSB galaxies constitute the $7.2\pm0.2\%$ of  galaxies in low-z cluster, when a cut at  1.2 $R_{200}$ was imposed. Their fraction increases from the outskirts toward cores. Considering the cluster  X-ray luminosity and velocity dispersion, they also found that the PSB fraction increases from the least toward the most massive and luminous clusters. 

\section{Galaxy properties and sample selection}\label{sec:gal_prop}
{ In the PM2GC,} 
we convert the observed SDSS photometry into rest-frame absolute magnitudes running INTERREST
\citep{Taylor2009}. 
This code is based on a number of template spectra and carries
out an interpolation from the observed photometry in bracketing
bands \citep[see][]{Rudnick2003}. We then derived rest-frame colors from the interpolated rest-frame  magnitudes. 
{ In WINGS+OmegaWINGS, rest-frame B and V absolute magnitudes are computed from  the observed ones \citep{Varela2009,Gullieusik2015} running the spectrophotometric model SINOPSIS \citep[SImulatiNg OPtical Spectra wIth Stellar population models,][]{Fritz2007, Fritz2011,Fritz2014}. The code is based on a stellar population synthesis technique that reproduces the observed optical galaxy spectra.} 

As presented by \cite{Poggianti2013,Paccagnella2017}, {  for both samples} we exploit SINOPSIS also to  obtain estimates of stellar masses and galaxy equivalent widths (EWs) for the most prominent spectral lines, both  in emission and in absorption. %The code is based on a stellar population synthesis technique that reproduces the observed optical galaxy spectra.
%All the main spectrophotometric features are reproduced by summing the theoretical spectra of simple stellar populations of 12 different ages (from 3 $\times 10^6$ to approximately 14$\times 10^9$ years).%}
We consider the stellar mass value corresponding to the mass locked into stars, considering both those that are  in the nuclear-burning phase, and remnants.

Observed EWs will be converted in rest frame values. We will adopt the usual convention of identifying absorption lines with positive values of the EWs and emission lines with negative ones. { \cite{Fritz2007} found that SINOPSIS is able to reliably measure emission and absorption lines in spectra with a S/N per pixel computed across the entire spectral range > 3. If this is not the case, SINOPSIS can misjudge noise as an emission. This happens especially in a spectral range where the lines of the Balmer series (in absorption) crowd, such as for the [O II] emission line.  Therefore, a peak between two such absorption lines can be misinterpreted as an [O II] emission. Simulations from the \cite{Fritz2007} have shown that a fair threshold value to get a reliable [O II] detection is -2 \AA{}. Lines with an EW higher (in absolute value) than this, are properly measured. }

We will use the morphological classification performed by MORPHOT \citep{Fasano2012}, an automatic tool designed to reproduce the visual classifications. %MORPHOT is based on the classical CAS (concentration/asymmetry/clumpiness) parameters, but introduces a set of additional indicators derived from digital imaging of galaxies. 
The code was applied to the B-band MGC images and V-band OmegaCAM images to identify ellipticals, lenticulars (S0s), and later-type galaxies (\citealt{Calvi2012}, Fasano et al. in prep).

The projected local galaxy density was derived by \citet{Vulcani2012} for each galaxy in the PM2GC survey from the circular area A that, in projection on the sky, encloses the 5th nearest projected neighbours within $\pm 1000 km/s$
brighter than M$_V=-19.85$, which is the V absolute magnitude limit at which the sample is spectroscopically complete.
The projected density is then $\Sigma = N/A$ in number of galaxies per Mpc$^2$. 

%\bbv{cosa fare per WINGS?}

\subsection{The spectral classification}\label{sec:spec_class}

\begin{figure*}
\begin{center}
 \includegraphics[scale=0.45, clip, trim=5 250 20 0]{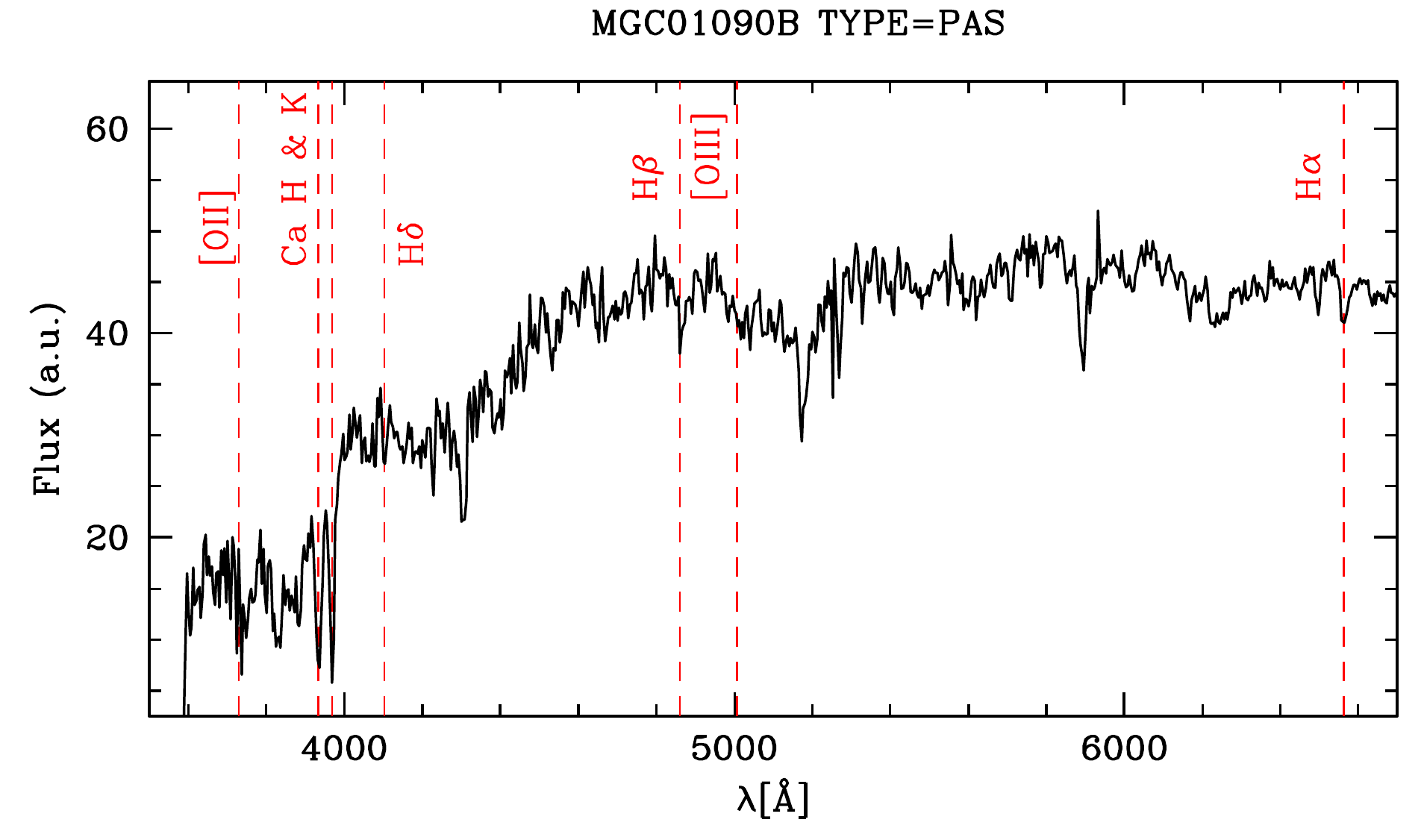}
 \includegraphics[scale=0.45, clip, trim=5 250 20 0]{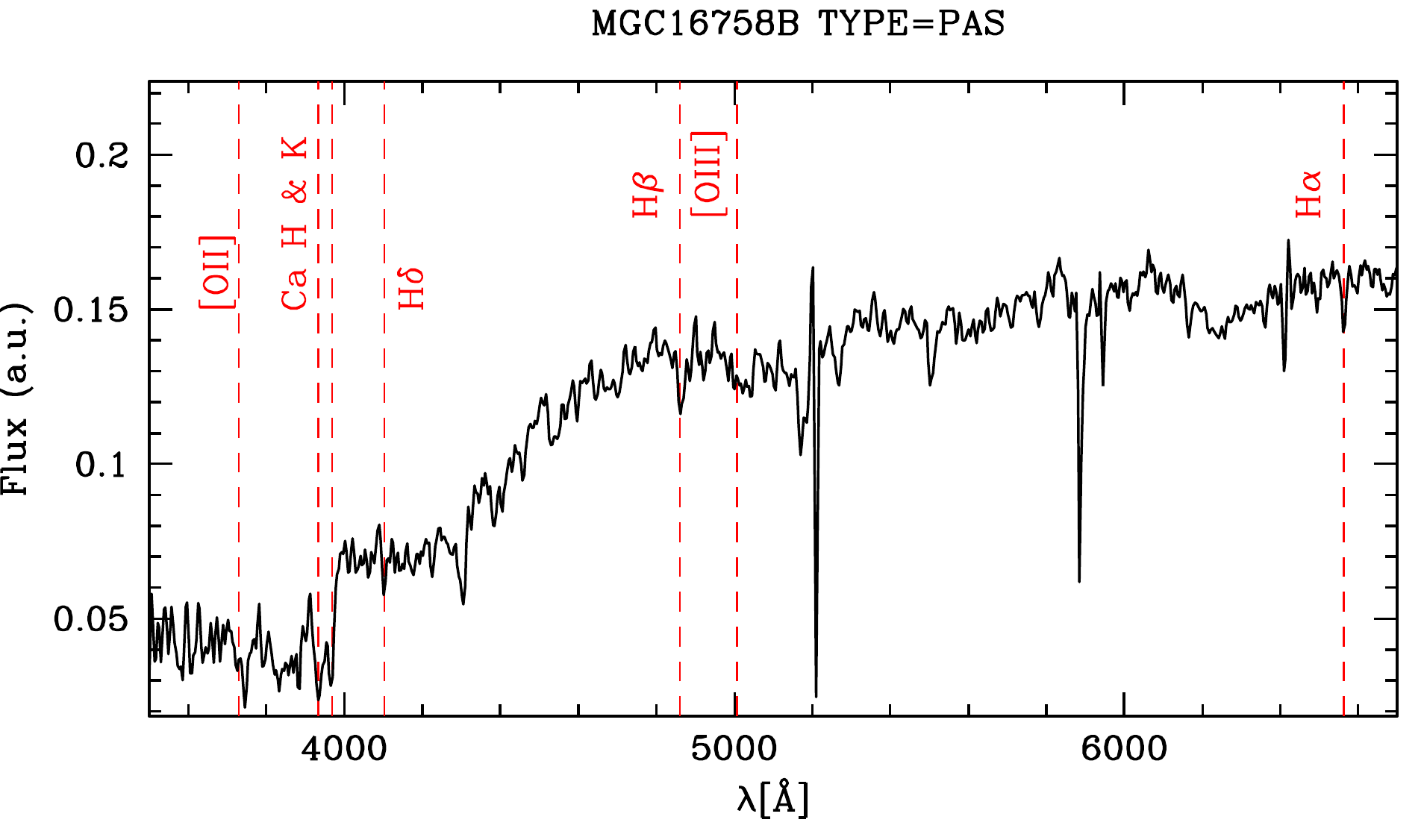}
 \includegraphics[scale=0.45, clip, trim=5 250 20 0]{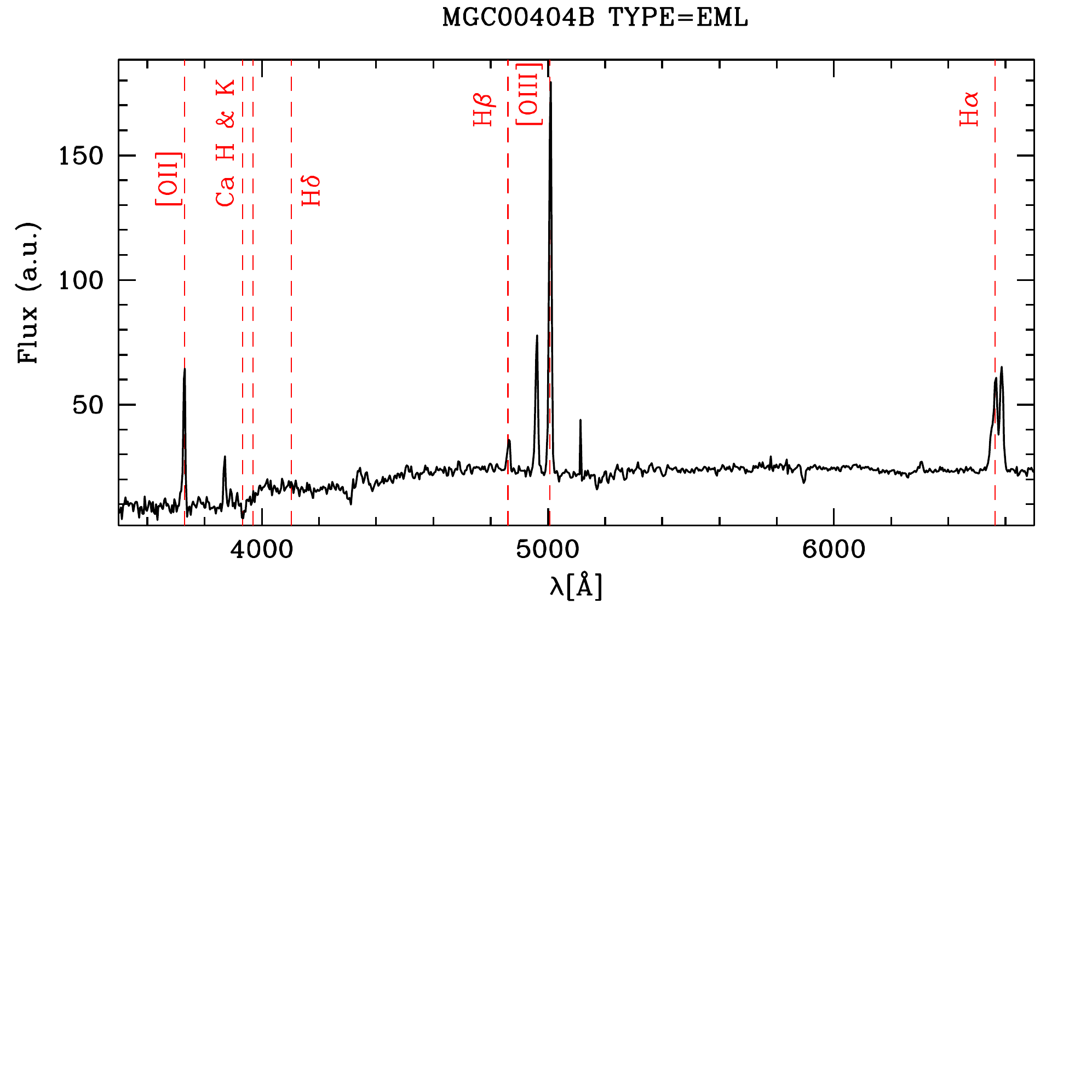}
 \includegraphics[scale=0.45, clip, trim=5 250 20 0]{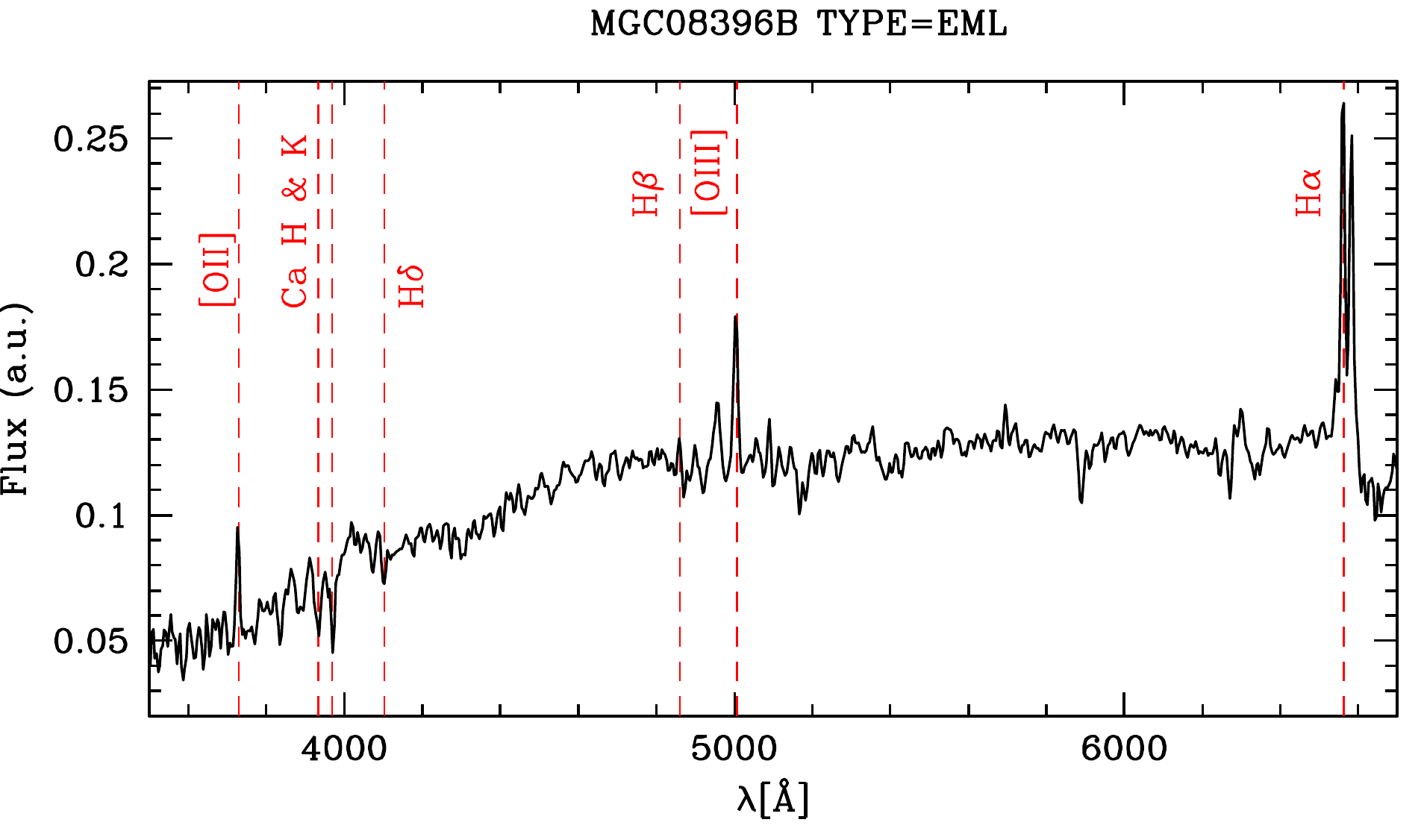}
 \includegraphics[scale=0.45, clip, trim=5 250 20 0]{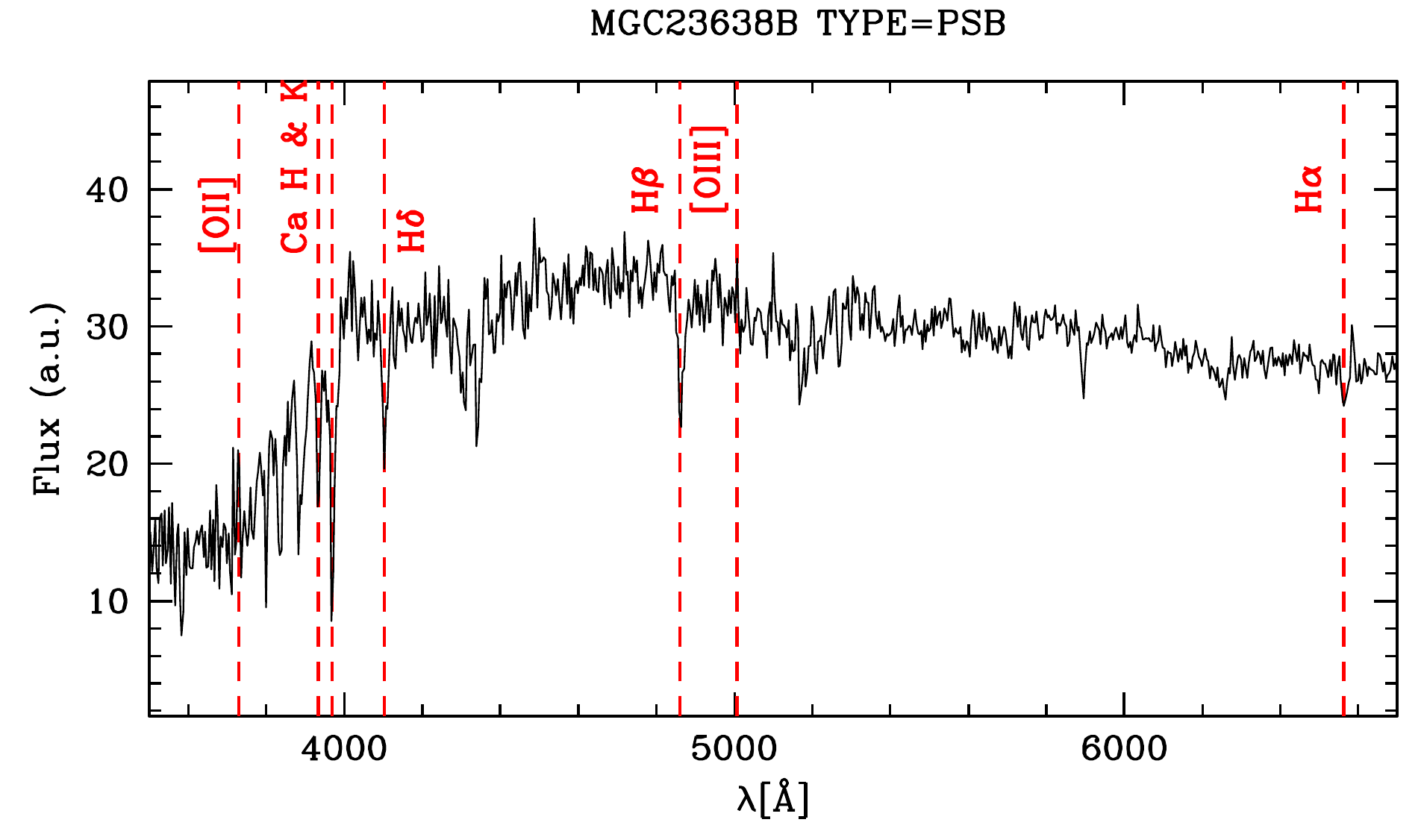}
 \includegraphics[scale=0.45, clip, trim=5 250 20 0]{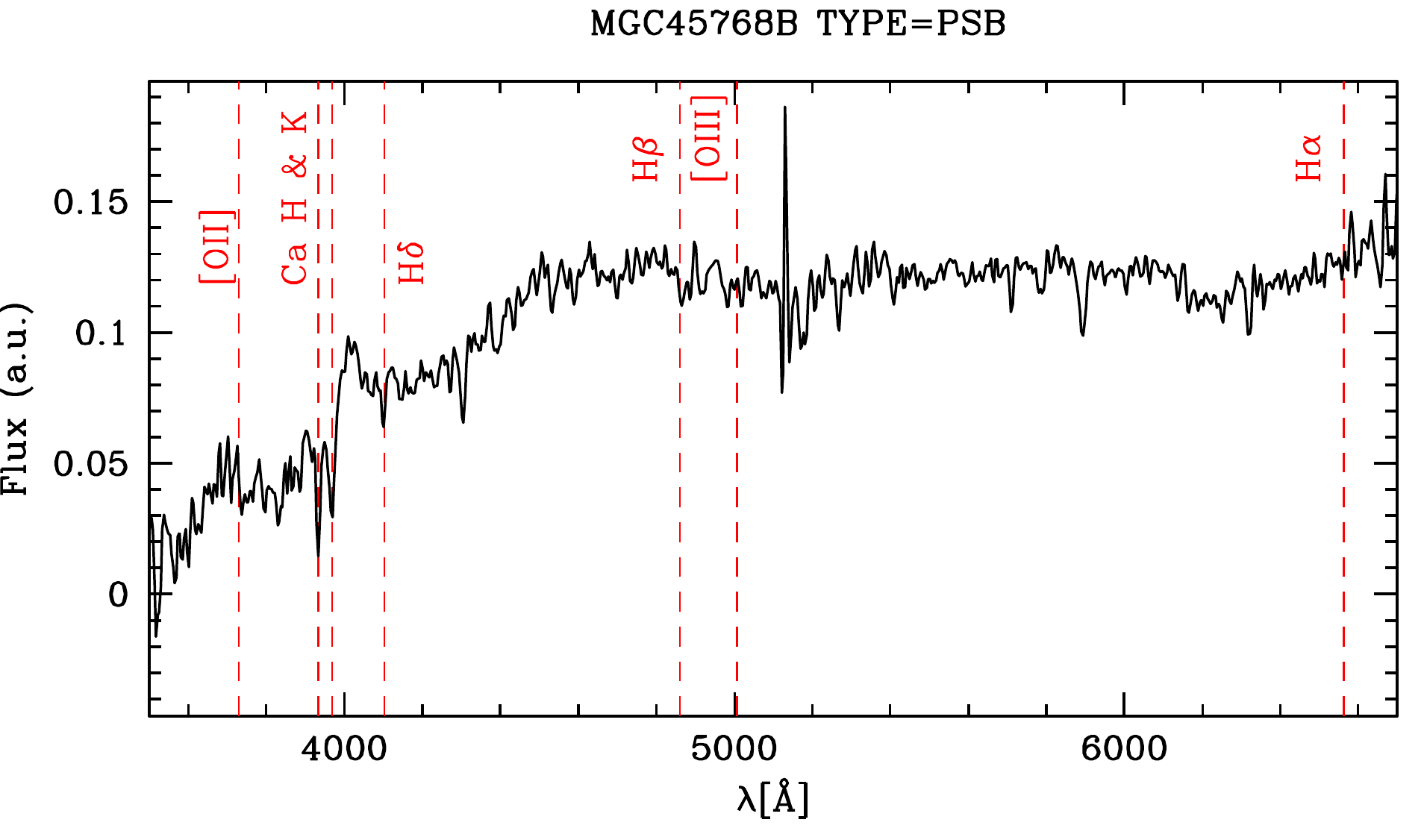}

\end{center}
\caption{{ Examples of spectra of PAS, PSB and EML galaxies in our samples are shown. Left panels show SDSS spectra, right panels show 2dF spectra.} \label{fig:spectra} }
\end{figure*}
 
%\bbv{add example of spectra}
We follow the same classification scheme presented in \cite{Paccagnella2017}, based on the measurement of the rest frame EWs. {We stress that SINOPSIS measures ``global'' equivalent widths, without explicitly separating the emission from the absorption component. The classification is therefore based on the actual measurements, without removing possible contamination of emission lines superimposed to the absorption lines.}  Briefly, we use  the rest frame EWs of [OII], H$\beta$, [OIII], H$\alpha$ in emission and H$\delta$ in absorption to subdivide the sample into the following three classes.
\begin{itemize}
\item Emission Line Galaxies (EML): galaxies whose spectra show any line in emission. {  These galaxies  have $|EW/err(EW)|\geq 3$ for at least one line present in the spectrum;}
\item Passive Galaxies (PAS): galaxies whose spectra show no emission lines  and weak H$\delta$ in absorption (EW(H$\delta$)<3 \AA). These galaxies are typically characterized by { spectra resembling those of K-type stars, lacking emission lines and with weak H$\delta$ in absorption, ($k$ spectra),} which is typical of passively evolving galaxies with neither current nor recent star formation activity normally found in passively evolving elliptical galaxies; %, resembling those of K-type stars; 
\item Post Starburst Galaxies (PSB): galaxies whose spectra show no emission lines  and strong H$\delta$ in absorption. These galaxies are typically characterized by $k+a$ ($3<$ EW(H$\delta)<8 \AA$) and $a+k$  (EW(H$\delta$)>8 $\AA$) spectra, and display a combination of signatures typical of both K and A-type stars.
\end{itemize}

Before proceeding with the analysis, we visually inspected all the spectra where only one emission line was detected and changed the galaxy spectral type where necessary. 
We also visually inspected all the PSB candidates and 
remeasured the H${\delta}$ EW for those  with an automatic measure higher than 5 \AA{} or with a S/N$\sim$1.
 { Note that our classification scheme assumes that any emission line would be coming from star-forming regions. If this assumption is not correct - for instance, when the contribution of [O II] from AGNs is significant - the PSB sample would be incomplete because AGNs would be mistakenly discarded as star-forming galaxies \citep[e.g.][]{Yan2006, Lemaux2017}.}
{ Figure \ref{fig:spectra} shows some spectra of galaxies in the PM2GC sample. Both typical SDSS and 2dF spectra for EML, PSB and PAS galaxies are shown.}

\subsection{The sample selection}
{ For the PM2GC, }
in the redshift range $0.03<z<0.11$, we consider all galaxies entering the sample with $M_B\leq -18.7$, which is the survey magnitude completeness limit. We  use galaxies in the three different environments: groups, binary and single systems. This sample includes 2642 objects. 

{ For the WINGS+OmegaWINGS sample, we consider the redshift range $0.04<z<0.07$ and include in the analysis only the 32 clusters with a global spectroscopic completeness higher than $\sim$50\% \citep{Paccagnella2017}. Cluster members within 1.2$R_{200}$ from the Brightest Cluster Galaxy enter the final sample. 
While in \cite{Paccagnella2017} we selected galaxies with a V magnitude brighter than 20, here, to directly compare the different data sets, we  apply the same magnitude cut of the PM2GC ($M_B=-18.7$). The main results of \cite{Paccagnella2017} still hold with the current selection.}

Note that { in both samples} we neglect the  contribution from AGN in our analysis. \citet{Alatalo2016} showed that this assumption  could in principle bias the results. 
Indeed, galaxies in which emission lines are are excited by the AGN mechanism or shocks rather than by  star formation, are excluded from the PSB sample. 
Nonetheless, to investigate the incidence of AGNs in { the PM2GC} sample, we have matched it with an AGN catalog from SDSS,\footnote{\url{ http://www.sdss3.org/dr10/spectro/spectro_access.php}} { based on emission line ratios,} finding that overall 51 galaxies ($2.0 \pm 0.3 \%$) that enter our selection are most likely hosting an AGN.
{ For the cluster sample, \cite{Guglielmo2015} found that in  WINGS  the AGN contribution in the star-forming galaxy population in a mass-limited sample  is approximately 1.6\% and \citep{Paccagnella2017} estimated that this fraction does not  change in the entire WINGS+OmegaWINGS sample.
}
Thus, including these galaxies in the EML population should not considerably affect the results.

\section{Results}
 \begin{table*}
\centering
\caption{Number and percentage of the different spectral types in the general field and its finer environments. Errors are binomial. The cluster sample is drawn from \citet{Paccagnella2017} and   will be discussed in Sec. \ref{sec:psb_envs}. Numbers are weighted for spectroscopic incompleteness, while raw numbers are given in brackets.\label{tab_frac2}} 
\begin{tabular}{ ll  |c c| c c| c c  }
\hline
\multicolumn{2}{c|}{\textbf{Environment}}  & \multicolumn{2}{c|}{\textbf{PAS}} & \multicolumn{2}{c|}{\textbf{PSB}} &  \multicolumn{2}{c}{\textbf{EML}}\\
& & N & \% & N & \% & N & \% \\
\hline
\multicolumn{2}{l|}{General field}  &   652 & 25$\pm$1 & 53 & 2.0$\pm$0.3 & 1937 & 73$\pm$1\\
&Groups &   373 & 36$\pm$1 & 31 &3.0$\pm$0.6 & 629 & 61$\pm$1 \\
&Real groups &   280 & 42$\pm$2 & 25 & 3.7$\pm$0.7 & 364 & 54$\pm$2 \\
&Binaries &   111 & 23$\pm$2 & 10 & 2.0$\pm$0.6 & 365 & 75$\pm$1\\
&Single galaxy &   168 & 15$\pm$1 & 12 & 1.1$\pm$0.4 & 943 & 84$\pm$1\\
\multicolumn{2}{l|}{Clusters}  &   3125 (1660) &60.9$\pm$0.6 & 229 (119) &4.5$\pm$0.3 & 1776 (1026) & 34.6$\pm$0.6\\
\hline
\end{tabular}
\end{table*}

%\bbv{aggiungere meglio confronti con i clusters?}

%\bbv{speigare meglio effetto fake groups}
\subsection{Properties of the different galaxy populations } %in the general field}

\begin{figure}
\begin{center}
\includegraphics[scale=0.33,clip, trim=0 10 130 20]{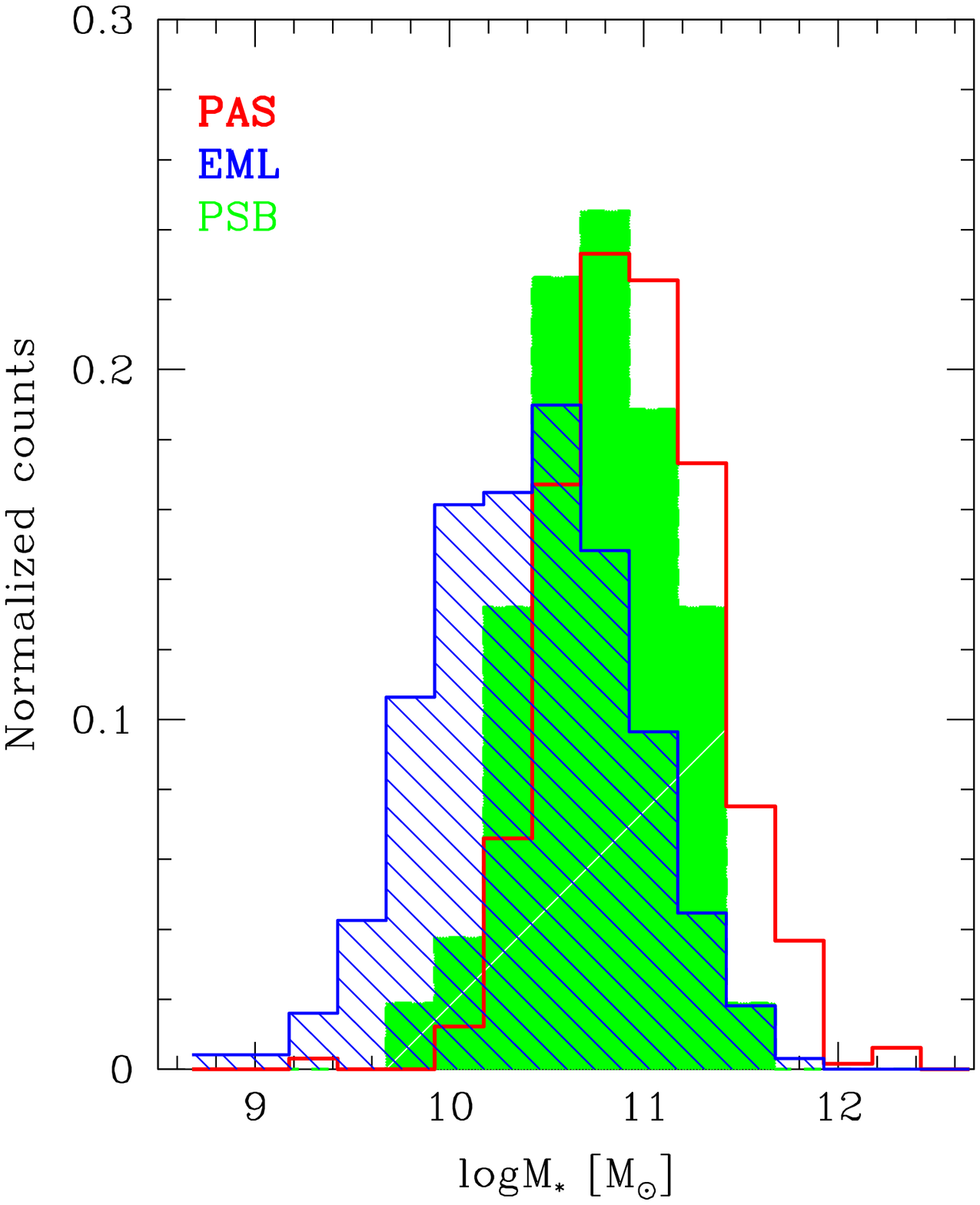}
\includegraphics[scale=0.33,clip, trim=70 10 240 10]{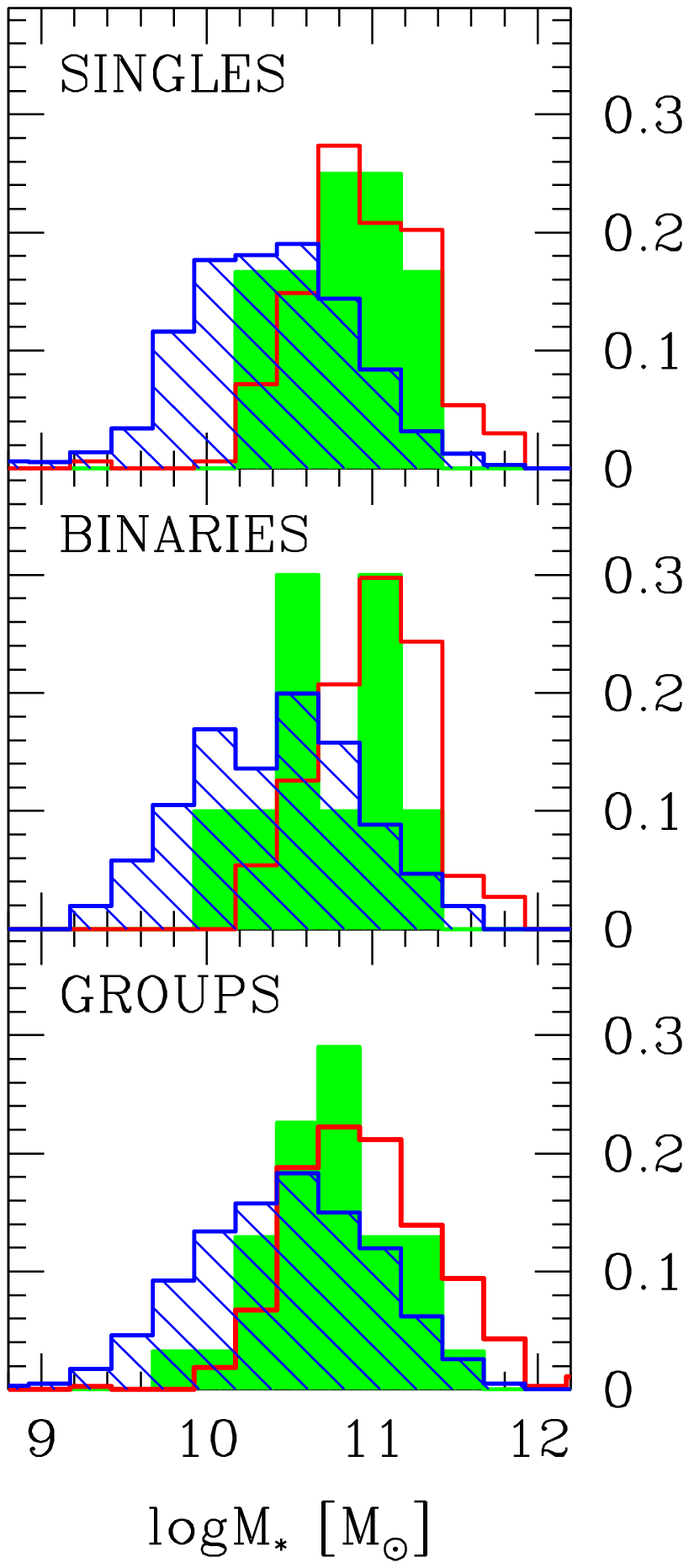}
\includegraphics[scale=0.33,clip,  trim=0 10 130 20]{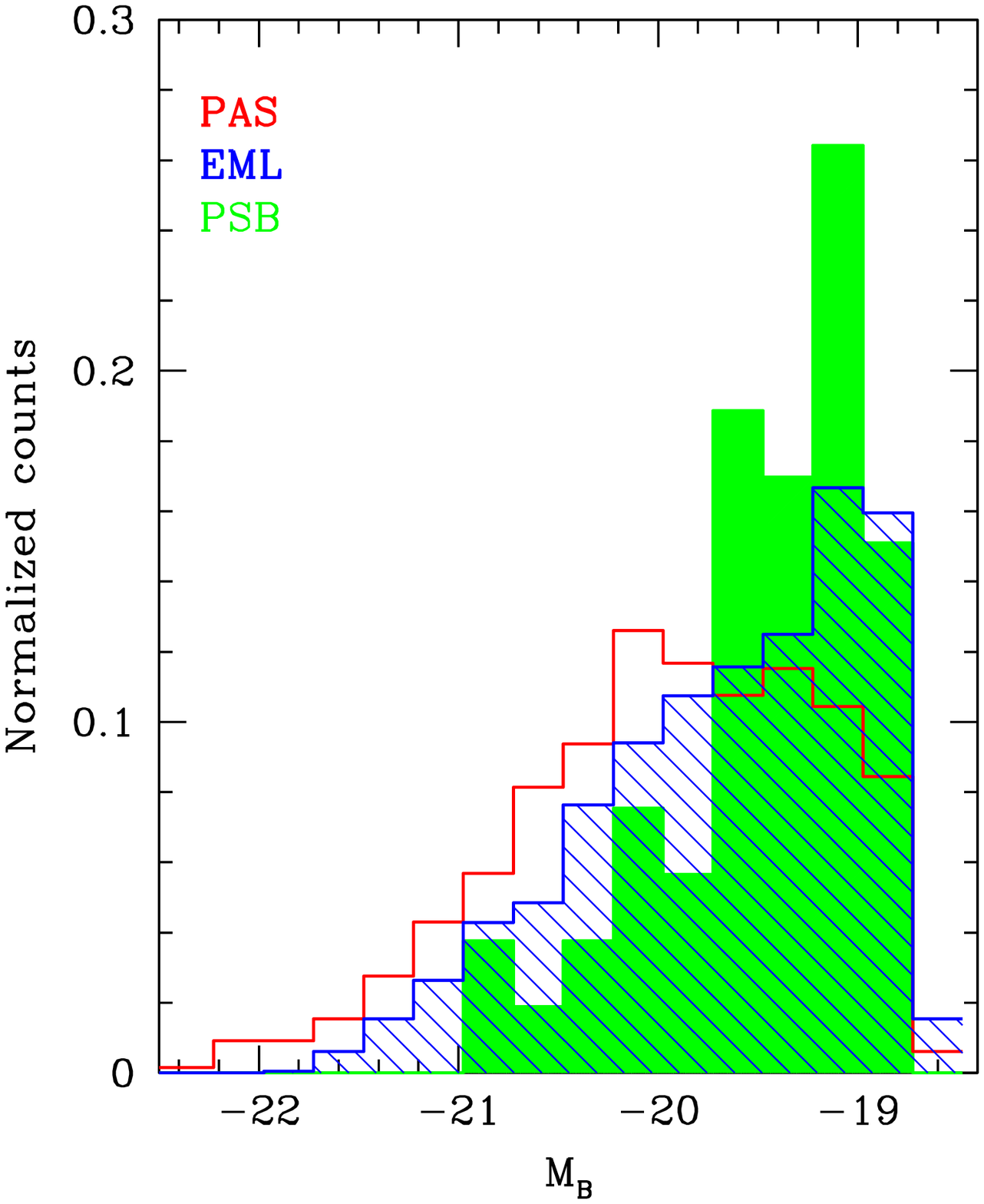}
\includegraphics[scale=0.33,clip, trim=170 10 150 10]{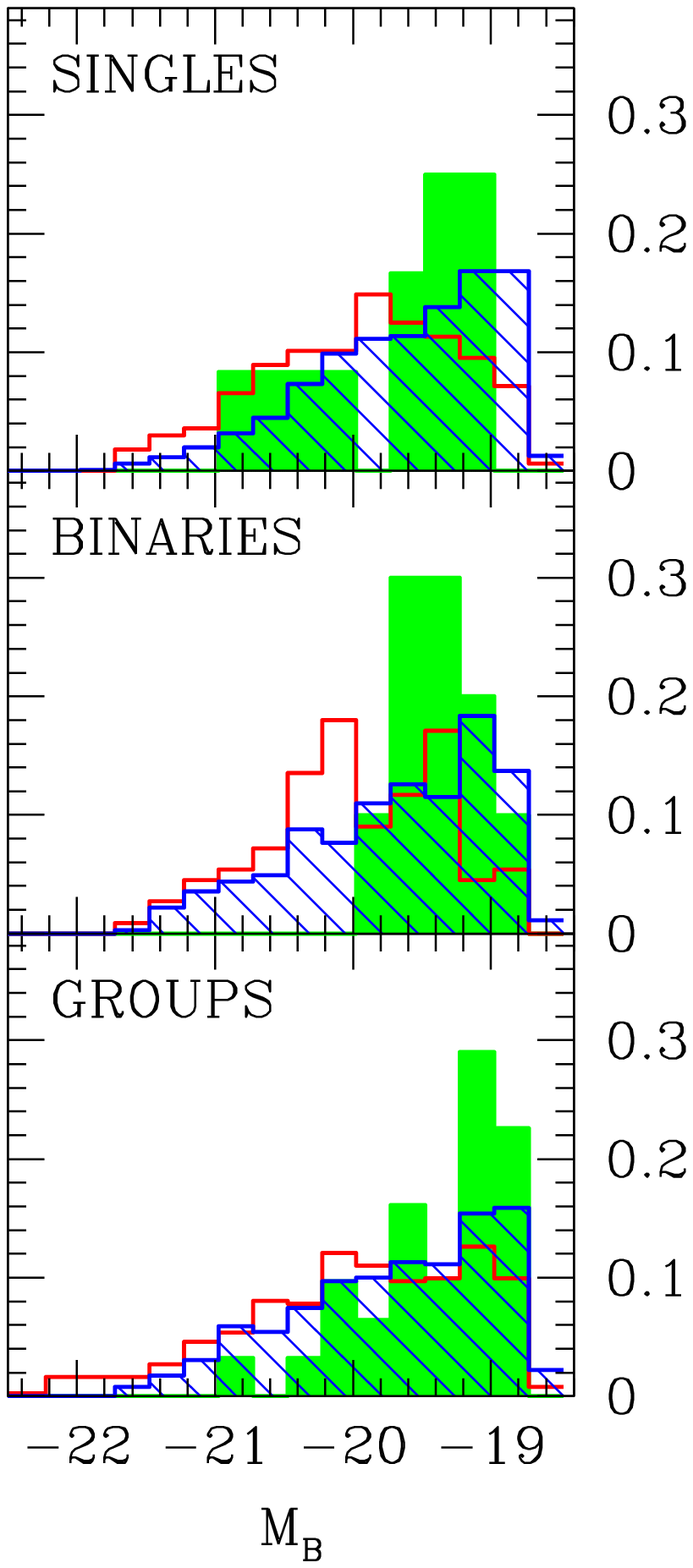}
\includegraphics[scale=0.33,clip,  trim=0 10 130 20]{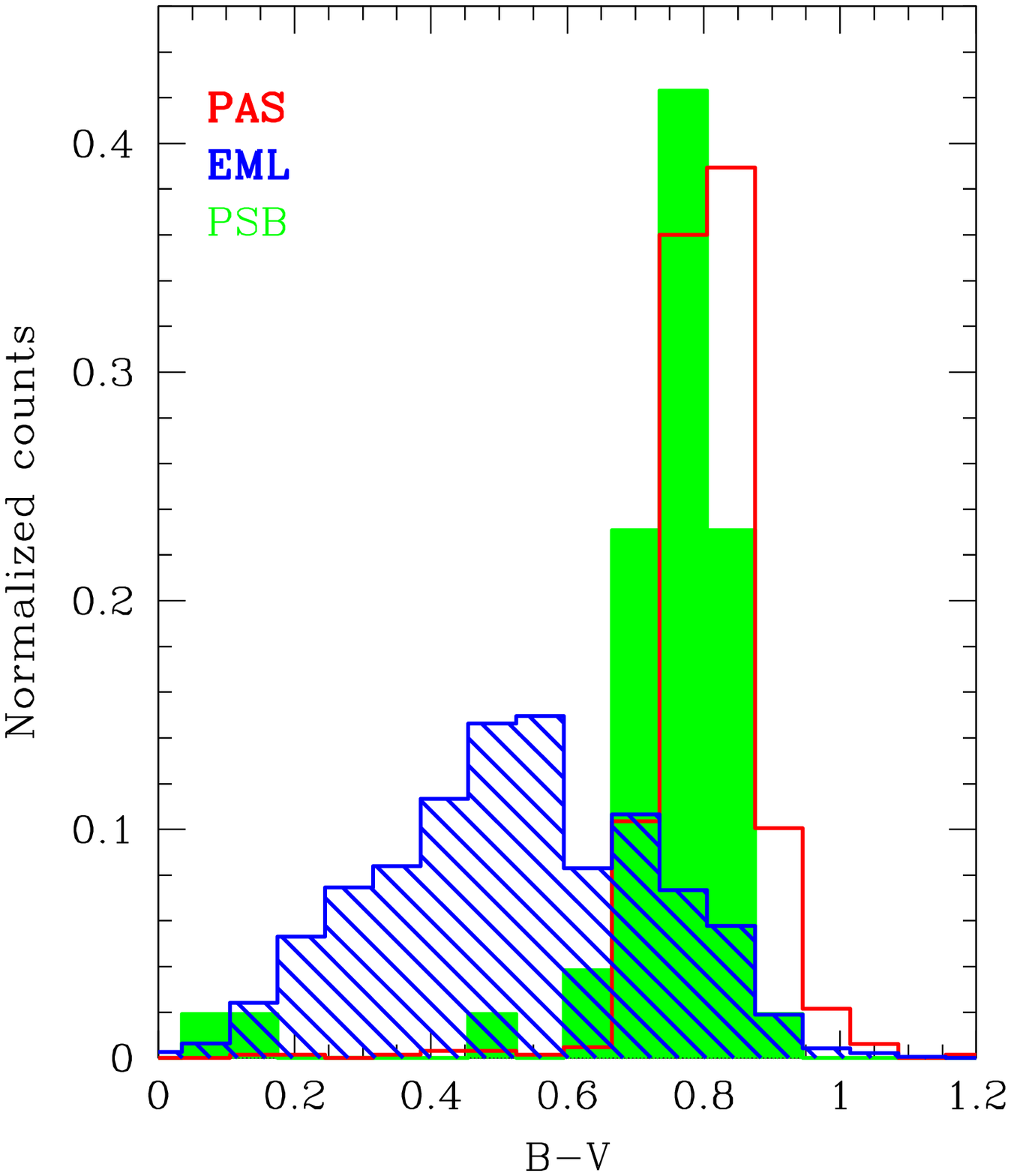}
\includegraphics[scale=0.33,clip, trim=170 10 150 10]{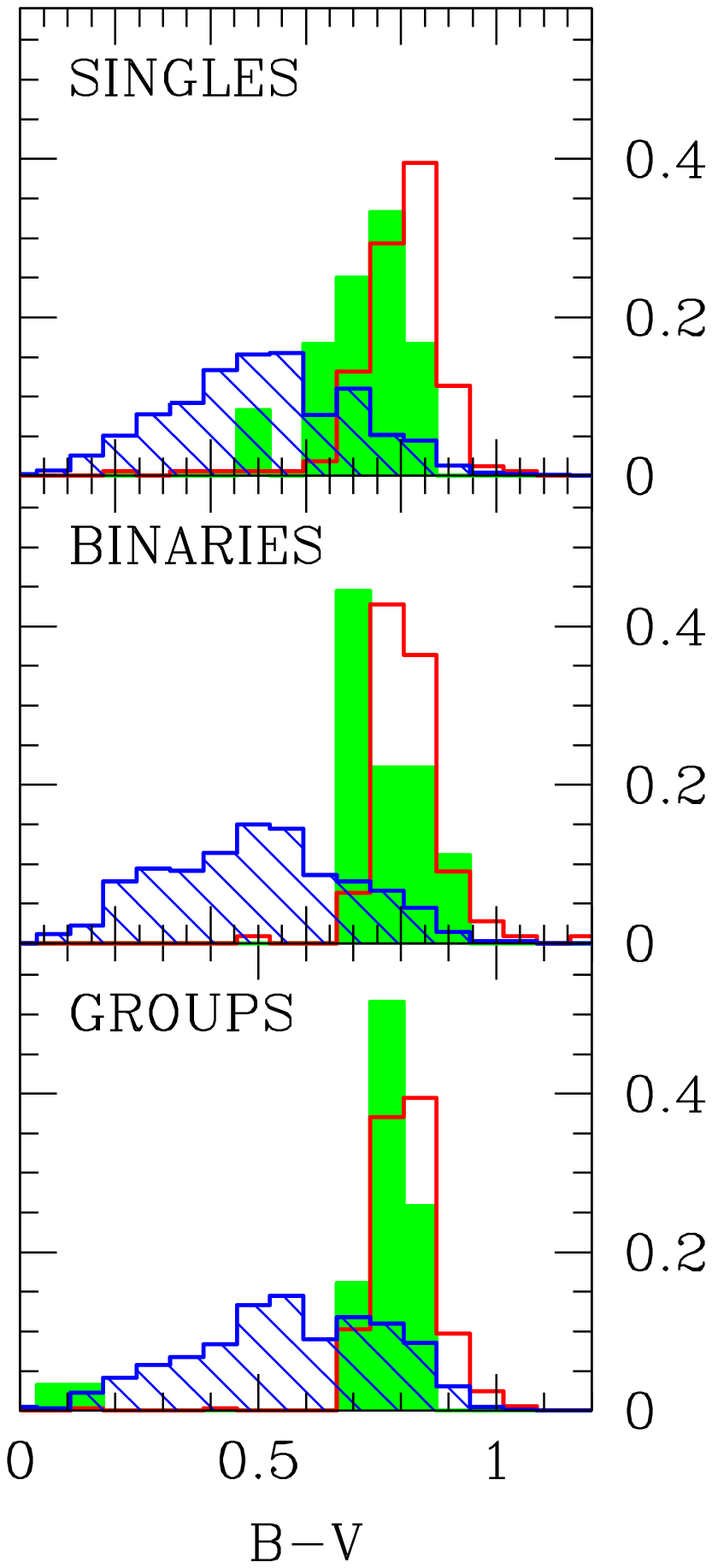}
\end{center}
\caption{Stellar mass (upper),  absolute B magnitude (central) and rest-frame (B-V) color (bottom) distribution for PAS (red, solid line histogram), EML (blue, hatched histogram) and PSB (green, filled histogram) galaxies. The main panel shows distribution in the general field, the subpanels show the distributions in the different environments.  Median values are reported in Tab.\ref{tab:mmdist_values}. \label{fig:mmdist} }
\end{figure}
 
We begin our analysis by examining the relative fractions of each spectral type in the different environments, as summarized in   Table \ref{tab_frac2}. In the general field, EML galaxies represent  $\sim73$\% of the whole galaxy population, PAS galaxies $\sim25\%$, while PSB galaxies are 2\% of the total.  
The incidence of each spectral type depends on environment, even though EMLs always dominate the population, followed by PASs and PSBs. 
Overall, the incidence of EMLs increases going from the more massive (i.e. groups) to the less massive (i.e. single) systems, while the incidence of PASs has the opposite behavior. 
These trends are even more outstanding if we consider only real groups. 
This result corroborates our distinction between fake and real groups: indeed fake groups are most likely just either binary or single systems that due to projection effect appear as group members, and are characterized by a higher fraction of EML galaxies. 
PSBs are always very rare, but in groups (both real and fake) they are more than twice as numerous as in single systems (3\% vs 1.1\%). In binaries they represent the 2\%. Therefore, the group environment seems to boost the presence of PSBs, confirming what previously found by \cite{Vulcani2015}. 
{ As results are robust against the inclusion of the fake groups, in what follows we will keep them in the analysis, so to have a better statistics.}

{ The trends already detected from the single systems to the groups extend to clusters: the fraction of PAS galaxies is much higher in clusters ($>60\%$) than in groups ($\sim40\%$), while the fraction of EML galaxies is much lower (35\% vs. 60\%).
The fraction of PSB galaxies grows from the poorest (only 1\% in single galaxies) to the richest environments (4.5\% in clusters).
}

%The cluster sample is drawn from \citet{Paccagnella2017} and will be discussed in Sec. \ref{sec:psb_envs}.

We define the ``active'' population (PSB+EML galaxies) as 
the population of galaxies that were star-forming $\lesssim$2 Gyr before the epoch of observation;  the PSB/active fraction can be therefore considered as a sort of quenching efficiency in truncating star formation in star-forming galaxies.
{ PSB galaxies 
represent $11.4\pm0.8\%$ of the active population in clusters \citep{Paccagnella2017},} %.PSB galaxies represent  
5$\pm$1\% of the active population in groups (6$\pm$2\% if only real groups are considered), and their incidence decreases in binaries (3$\pm$1\%) and single galaxies (1.3$\pm$0.6\%), suggesting that { in clusters and} groups the suppression of the star formation is more efficient than in other environments.  

Figure \ref{fig:mmdist} shows the stellar mass,  the B-band absolute magnitude  and the rest-frame (B-V) color distributions for the different spectral types both in the general field and in its finer environments{ , for galaxies in the magnitude limited sample}. %We show the results for all groups, but  we note that our findings do not change when using only real groups. 
The median values are given in Tab.\ref{tab:mmdist_values}. Errors are computed as $1.253\sigma /\sqrt{n}$ with $\sigma$ standard deviation and $n$ number of objects.
First focusing on the general field, EML and PAS galaxies have quite different mass/magnitude/color distributions, dominating respectively the low and high mass/luminosity and blue/red tails. 
 PSB galaxies are characterized by mass, luminosity and color distributions that are overall intermediate between that of the PAS and EML galaxies, lacking both the least and most massive objects and the reddest and the bluest ones.  No different trends emerge for galaxies in the finer environments and the median values of both stellar mass and magnitude of the different spectral type galaxies do not depend on environment.

{ Similarly, in clusters, PSB magnitudes, stellar masses, and  colors typically show an intermediate behavior between passive and emission line galaxies, indication of a population in transition from being star forming to passive \citep{Paccagnella2017}.}
The different mass distribution of the different %samples 
spectral types
is mainly driven by the fact that we are analyzing a magnitude limited sample. To obtain a mass complete sample, we should disregard galaxies with $\log(
M_\ast/M_\odot)$<10.44. Figure \ref{fig:mmdist} shows the the vast majority of galaxies below this threshold are EML galaxies. {This cut would however reduce the sample size by $\sim$40\%.} 
%As in what follows we will never study trends as a function of stellar mass, we are confident that our magnitude limited sample is representative of the local population of galaxies.} 

 \begin{table*}
\caption{Median values along with errors of stellar mass and absolute B magnitude for galaxies of different spectral type in the different environments.  \label{tab:mmdist_values}} 
\centering
\begin{tabular}{ l  |ccc| ccc| ccc | }
\hline
Environment & \multicolumn{3}{c|}{\textbf{PAS}} & \multicolumn{3}{c|}{\textbf{PSB}} &  \multicolumn{3}{c}{\textbf{EML}}\\
 & logM$_*$ & M$_B$ & (B-V)$_{\rm rf}$ & logM$_*$ & M$_B$ & (B-V)$_{\rm rf}$& logM$_*$ & M$_B$ & (B-V)$_{\rm rf}$\\
\hline
General field &   10.94$\pm$0.02 & -19.91$\pm$0.04 & 0.81$\pm$0.01 &10.76$\pm$0.06 & -19.28$\pm$0.09 &0.77$\pm$0.03 &10.42$\pm$0.02 & -19.54$\pm$0.02 & 0.52$\pm$0.01\\
Groups &   10.93$\pm$0.03 & -19.90$\pm$0.05 & 0.81$\pm$0.01 &10.76$\pm$0.09 & -19.21$\pm$0.01 & 0.78$\pm$0.04 &10.48$\pm$0.03 & -19.61$\pm$0.04 &0.55$\pm$0.01\\
Binaries &   11.01$\pm$0.04 & -20.01$\pm$0.08 & 0.81$\pm$0.01 &10.76$\pm$0.15 & -19.25$\pm$0.11 & 0.8$\pm$0.1&10.44$\pm$0.04 & -19.59$\pm$0.05 & 0.51$\pm$0.01\\
Singles &   10.92$\pm$0.04 & -19.87$\pm$0.07 & 0.81$\pm$0.01 &10.91$\pm$0.11 & -19.47$\pm$0.22 & 0.76$\pm$0.03 &10.38$\pm$0.02 & -19.50$\pm$0.02 & 0.51$\pm$0.01\\
\hline
\end{tabular}
\end{table*}
 
\begin{table*}
\centering
\caption{Morphological fractions for  galaxies of different spectral type in the different environments. Errors are binomial. \label{tab:morf} }
\begin{tabular}{l|ccc|ccc|ccc}
\hline
 & \multicolumn{3}{c|}{{ GROUPS}} & \multicolumn{3}{c|}{{ BINARIES}} & \multicolumn{3}{c}{{ SINGLES}}\\
\hline
  \multicolumn{1}{c}{} &
  \multicolumn{1}{c|}{E} &
  \multicolumn{1}{|c}{S0} &
  \multicolumn{1}{c}{LATE} &
  \multicolumn{1}{c|}{E} &
  \multicolumn{1}{c}{S0} &
  \multicolumn{1}{c}{LATE} &
  \multicolumn{1}{c}{E} &
  \multicolumn{1}{c}{S0} &
  \multicolumn{1}{c}{LATE}\\
\hline
  PAS & 0.42$\pm$0.03 & 0.43$\pm$0.03 & 0.15$\pm$0.02 & 0.39$\pm$0.05 & 0.41$\pm$0.05 & 0.20$\pm$0.04 & 0.39$\pm$0.05 &  0.39$\pm$0.03 & 0.22$\pm$0.03 \\
  PSB & 0.3$\pm$0.1   & 0.4$\pm$0.1   & 0.3$\pm$0.1   & 0.3$\pm$0.2   & 0.7$\pm$0.2   & 0.0$\pm$0.2   & 0.3$\pm$0.3   &  0.4$\pm$0.2   & 0.4$\pm$0.2\\
  EML & 0.08$\pm$0.01 & 0.18$\pm$0.02 & 0.74$\pm$0.02 & 0.07$\pm$0.01 & 0.14$\pm$0.02 & 0.79$\pm$0.02 & 0.09$\pm$0.01 &  0.14$\pm$0.01 & 0.77$\pm$0.01 \\
\hline\end{tabular}
\end{table*}

\subsubsection{Morphologies of the different galaxy populations}

\begin{figure}
%\vspace{-110pt}
\includegraphics[scale=0.44,clip, trim=2 25 20 290]{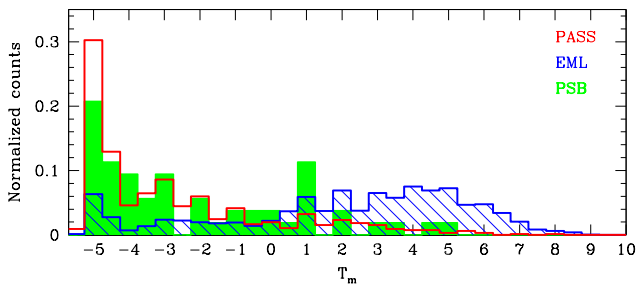}
\caption{{ Fine morphological distribution of galaxies of the different spectral types for galaxies in the general field.} \label{fig:morf}  }
\end{figure}

%\bbv{aggiungere figura distribuzioni}
The analysis of the morphologies of the different galaxy populations can shed light on their formation. 
{ Figure \ref{fig:morf} shows the fine morphological distribution of galaxies in the general field. PAS galaxies are clearly skewed towards earlier morphologies, mainly ellipticals and S0s. PSBs include a wide range of morphologies, ranging from ellipticals to intermediate spirals. Most of EMLs are intermediate and late-type spirals, even though few objects show an early type morphology (as discussed in \citealt{Vulcani2015}).}

Table \ref{tab:morf} presents the  morphological distribution of galaxies of the different spectral types, in the different environments. As expected, PAS, EML and PSB galaxies have different morphologies. However,  there is not a one-to-one correspondence between spectral type and morphology, suggesting that in many cases star-formation properties and galaxy appearance must change on different timescales, in agreement with previous studies \citep[e.g.,][]{Dressler1999, Fritz2014, Vulcani2015, Pawlik2016}{ , or that these galaxies are actually  rejuvenating and about to return to the blue sequence \citep[e.g.][]{Dressler2013, Abramson2013, Rowlands2015, French2015, Wild2016}. }

In all the environments, 74-79\% of EML galaxies present a late-type morphology, 14-18\% of them are classified as S0s, less than 10\% as ellipticals.  These fractions suggest that the morphology of EMLs  does not vary with
the environment. Similarly, { within the errors,} %aided by the large uncertainties,
also the morphologies of PSB galaxies seem to not depend on the environment: $\sim$30\% of them are ellipticals, $\sim 40-70\%$ are S0s, 0-40\% are late types. { Similar fractions are also found in local clusters \citep{Paccagnella2017}.}
Finally, the environment has also little influence on PAS galaxies: $\sim$15-20\% of them are late types, while the others are evenly split between S0 and ellipticals. There is a hint that in groups the fraction of late types is smaller, suggesting that this environment  accelerates morphological transformations after the shut off of the star formation, but the evidence in only marginal ($<2\sigma$, with $\sigma$ binomial error on the fractions).

\subsection{Properties of the different galaxy populations at different local densities in the general field}
%\bbv{discuss KS}
The local density is a parametrization of the environment based on the number of nearby galaxy companions, therefore it should better trace processes like harassment and galaxy-galaxy interactions. If, as discussed in Sec.\ref{sec:intro}, 
field post-starburst galaxies are indeed  associated with galaxy interactions and mergers \citep{Bekki2001, Quintero2004, Blake2004,Goto2005, Hogg2006, Mahajan2013}, we should find them at relatively  higher local densities. {Indeed, \cite{Kampczyk2013} found that close kinematic pairs  preferentially reside in overdense environments and, since the merger timescales is only marginally dependent of environment,  this translates to a three times higher merger rate}.

\begin{figure}
\centering
%\vspace{-60pt}
\includegraphics[scale=0.53,clip, trim=3 10 10 10]{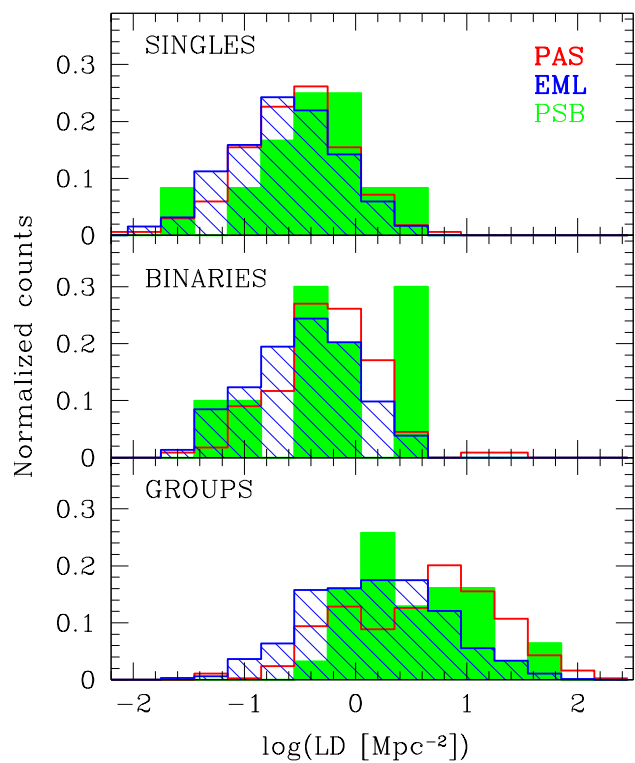}
\caption{Projected local density distribution for galaxies of the different spectral types (PAS: red, PSB: green, EML: blues), in the different environments (singles: upper panel, binaries: middle panel, groups: bottom panel). \label{fig:LD}}
\end{figure}

Figure \ref{fig:LD} shows the distribution of the projected local density for galaxies of the different spectral types in the different environments. Median values along with errors are provided in Tab.\ref{tab:ld}.
As expected, galaxies in singles and binaries are shifted toward lower density values than galaxies in groups, and in any given environment EML galaxies tend to be found at lower density values than PAS galaxies.  
Trends for PSBs are more interesting. { In binary and singles,} the distribution of PSBs is skewed toward larger local density values than the other two  populations,  supporting the scenario according to which PSBs are the result of galaxy interactions.
%Note, however, that  in binaries PSBs also form a tail at low density values, where the merging scenario does not hold.\ap{da rivedere questa parte visto il nuovo plot delle ld e le nuove mediane }
{ To probe the results on a more solid statistical ground, we performed Kolmogorov-Smirnov  (K-S) tests between the different samples. As expected, in all environments, PAS and EML galaxies are always significantly different (P$_{K-S}<<0.05$). As far as PSBs in concerned, they are significantly different from EMLs in groups and single systems, while the K-S is not able to robustly determine deviations in binaries, maybe due to the low number statistics. Finally, the K-S is not able to conclude that PSBs and PASs are drawn from different distributions. 
These results therefore suggest that  EMLs in high density environments most likely quench, turning into PSBs and eventually PASs. 
}
%\bbv{pensare a cosa vuol dire}
%\bmp{il KS non funziona cosi', parlare}

\begin{table}
\centering
\caption{Median local density values along with errors for  galaxies of different spectral type in the different environments. \label{tab:ld} }
\centering
\begin{tabular}{ l  ccc }
\hline
&  {\textbf{PAS}} &{\textbf{PSB}} & {\textbf{EML}}\\
Groups & 0.69$\pm$0.05 & 0.6$\pm$0.2 & 0.16$\pm$0.04  \\
Binaries & -0.3$\pm$0.1 & -0.2$\pm$0.2 & -0.46$\pm$0.03\\
Singles & -0.51$\pm$0.06 & -0.3$\pm$0.1 & -0.64$\pm$0.02\\
\hline
\end{tabular}
\end{table}

\subsection{Post-starburst galaxies as a function of halo masses}\label{sec:psb_envs}

%\bbv{commentare mass limited sample}
In the previous section we have investigated the incidence and the properties of galaxies of the different spectral types in the different environments, ranging from the single systems to the { clusters.}

\begin{figure}
\centering
\includegraphics[scale=0.43,clip, trim=0 55 10 225]{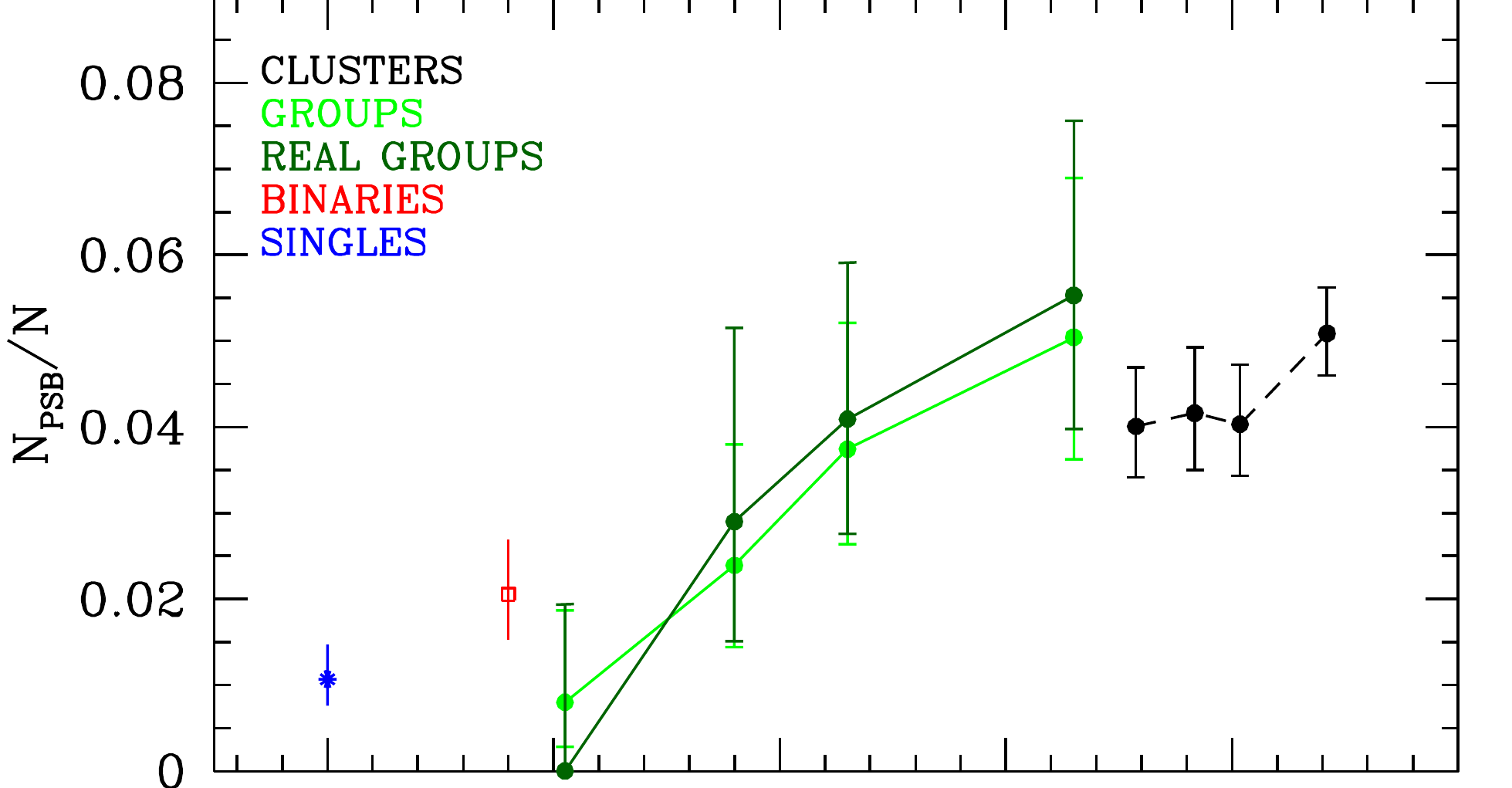}
\includegraphics[scale=0.43,clip, trim=0 0 10 220]{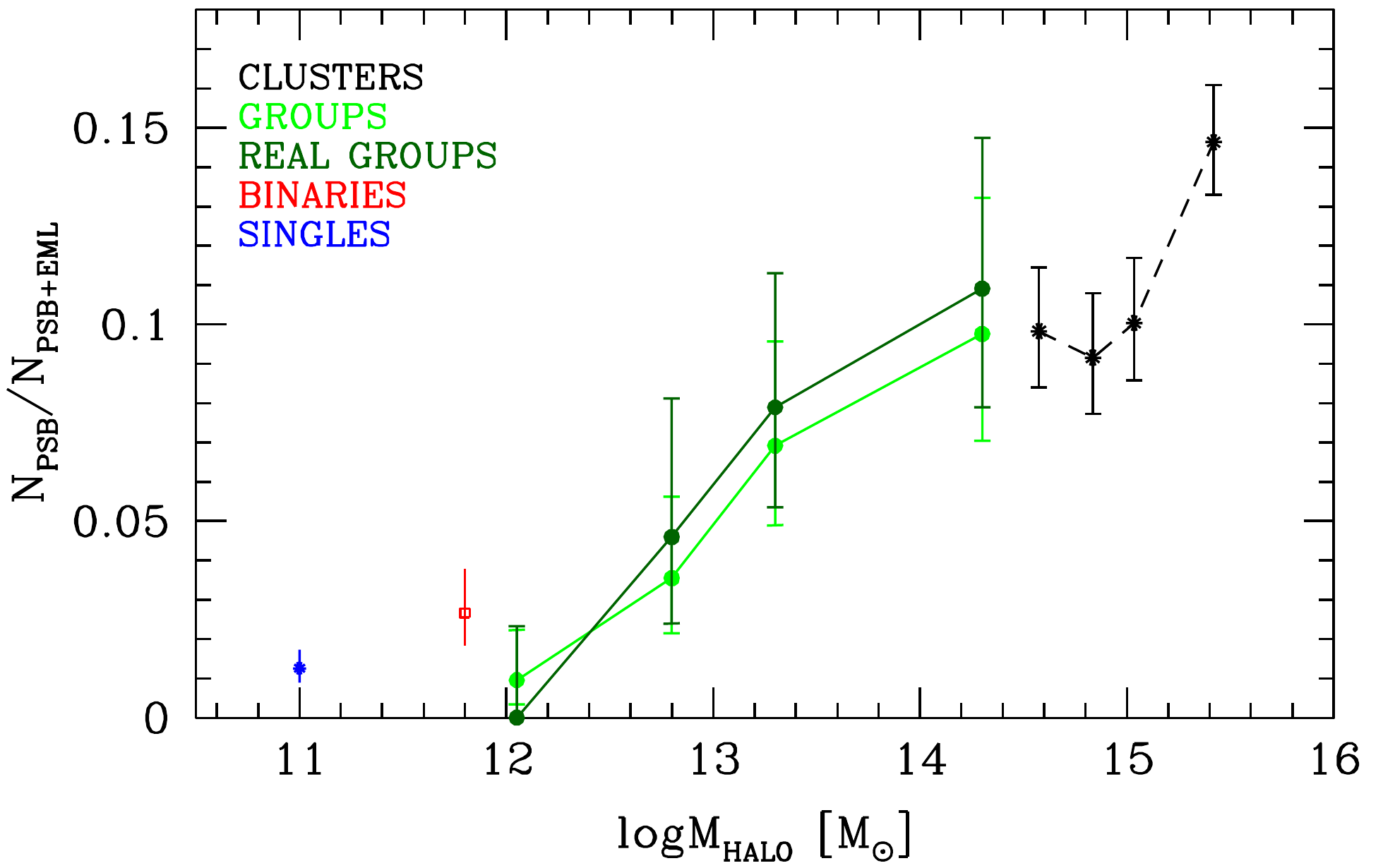}
\caption{Fraction of PSB galaxies with respect to the total population (upper panel) and with respect  to the active population (EML+PSB, lower panel) as a function of the halo mass.
Black dots and dashed lines represent the cluster data; green (dark green) dots and solid lines represent the PSB fractions in groups (real groups). The red empty square and the blue point at $\log (M_{halo}/M_\odot)=$11.8 and 11  show the value for binaries and single galaxies, respectively. Errors are binomial.\label{fraz_mhalo}}
\end{figure}

We are now in the position of studying the PSB population as a function of the mass of the dark matter halo.
Considering the different environments together, namely single galaxies, binaries, groups and clusters, the halo mass range covered goes from $10^{11}$ M$_{\odot}$ to above $10^{15}$ M$_{\odot}$. 
The upper panel of Fig.\ref{fraz_mhalo} shows the  fraction of PSBs as a function of the halo mass.
A clear trend emerges: the fraction of PSB galaxies increases with the halo mass.  
If only real groups are considered, this trend is even more pronounced.
Not only the incidence of PSB galaxies becomes more important going to higher halo masses,
but also the quenching efficiency (N$_{PSB}$/N$_{PSB+EML}$) significantly increases, as  shown in the lower panel of Fig. \ref{fraz_mhalo}.
It therefore appears  clear that the environment plays a major role in inducing the formation of PSB galaxies, which are more frequent in massive haloes, as discussed in the next Section. 
{ Note that these results still hold when applying a common cut in stellar mass in the two samples, implying that the different mass distribution can not be responsible for the observed trends. }

\section{Discussion}
In this work we have characterized the PSB population in the general field at $0.03<z<0.11$ and investigated their incidence as a function of environment.  { We also contrasted their properties to those of PSB galaxies in clusters at similar redshift \citep{Paccagnella2017}.}
Overall, PSBs have properties {(stellar mass, rest-frame B magnitude, B-V color)} intermediate between those of PAS and EML galaxies, suggesting that these objects are in a transitional stage between star-forming blue cloud galaxies and quiescent red sequence galaxies \citep[in agreement with e.g.][]{Caldwell1996, Zabludoff1996, Norton2001, Pracy2009,Zwaan2013, Yesuf2014, Pattarakijwanich2016, Paccagnella2017}.

{ Our main result is that the frequency of PSB galaxies strongly depends on halo mass.}
The quenching efficiency is 
enhanced in more massive systems: the ratio  of PSBs to PSBs+EMLs in clusters ($\sim11\%$) is twice the same ratio in groups ($\sim6\%$). The same ratio is halved in binary systems ($\sim3\%$) and further reduced among single galaxies ($\sim1\%$). 
These trends are even more outstanding when the fraction of PSBs is plotted as a function of halo mass (Fig. \ref{fraz_mhalo}).
Therefore, massive environments boost the presence of PSBs. 

These results are in agreement with those  presented by \citet{Gavazzi2010} for the PSB population within the Coma supercluster. They found that the PSB fraction  significantly increases  from the cosmic web to groups, to the cluster center.
These findings suggest that  any physical mechanism that produces PSB galaxies must depend in some way on the global environment. 

We remind the reader that the PSB signature in a galaxy spectrum is the consequence of a truncation of the star formation on time-scales as short as or shorter than $\sim$100 Myr, but there are no signs on the spectrum that allow us
to distinguish among the  different mechanisms responsible for the abrupt shut down of the star formation.

Different scenarios have been proposed to explain the formation of PSB galaxies in the different environments.
In clusters, the truncation of the star formation is most likely due to the interaction of the galaxy with the intracluster medium \citep[e.g.][]{Dressler1982,Couch1987,Poggianti2009,Paccagnella2017}. As already discussed and extensively modeled (see for example \citealt{Boselli2008} and references therein), ram pressure stripping could be the driving mechanism. This mechanism is expected to be strongest in the {core of} most massive clusters \citep[e.g.][]{Gunn1972, Jaffe2015, Zinger2018, Hwang2018}, and this can explain the observed trend with halo mass. Trends in groups can be explained as the natural sequence of intergalactic medium (IGM), with more massive systems having a denser IGM.

When smaller systems as { small groups}, binaries or  single galaxies are included in the global picture, other processes might need to be considered. In these environments, the most obvious candidate mechanisms are close galaxy interactions or mergers. 
According to numerical simulations \citep[e.g.][]{Barnes1991, Barnes1996, Bekki2001, Bekki2005, Hopkins2006, Hopkins2008, Snyder2011}, gas-rich major mergers can induce strong starbursts   that quickly exhaust a substantial amount of the gas reservoir \citep{DiMatteo2005}, and can transform late-type galaxies into early types both structurally and kinematically \citep{Toomre1972, Naab2003}.
Many works support this scenario, from both the theoretical and observational point of view \citep[][and references therein]{Bekki2001,Goto2005,Blake2004}. Theoretical simulations indeed indicate that the PSB spectrum marks a late stage of the merging event, when the merging companion is no longer identifiable as the morphological signatures of the merging.

The signs of morphological disturbance found in many PSB galaxies provides strong evidence sustaining this statement \citep{Zabludoff1996, Blake2004, Goto2005, Yang2008}. 
Galaxy-galaxy interactions are traced by  local density estimates and our analysis considering the projected local density (Fig.\ref{fig:LD}) shows that  the distribution of PSB is skewed toward larger local density values than the other two  populations,  supporting the scenario according to which PSB are the result of galaxy interactions. 

However, we did not detect any signs of disturbance in the morphology distribution of the PSBs in low mass halos, in agreement e.g. with \cite{Zabludoff1996, Blake2004, Yang2008}. These results suggest that possibly less disruptive events, like minor mergers and accretion, are at play \citep[see also][]{Dressler2013}. This conclusion is supported also by the fact that the redshift-evolution of the PSB number density found by \cite{Wild2009} is $\sim$100 times that of the major-merger rate estimated by \cite{deRavel2009}. 

Our results provide evidence that processes specific to the densest environments, such as ram pressure stripping, are responsible for a large fraction of PSB galaxies in dense environments. These processes act on a larger fraction of galaxies than alternative processes leading to PSB galaxies in the sparsest environments, such as galaxy interactions.

\section{Summary and Conclusions}\label{sec:conc}
\label{conclusion}
Exploiting  an absolute magnitude limited sample of galaxies ($M_B\leq -18.7$) at $0.03<z<0.11$ in the general field drawn from the PM2GC \citep{Calvi2011}, we have investigated the occurrence and the  properties of galaxies of different spectral types  in different environments.  
We used different parameterizations of the environment, in order to have a broad understanding of its  influence in shaping galaxy properties. First, we exploited the outputs of a FoF algorithm, which  subdivided galaxies into single, binary and group systems. Second, we used projected local density estimates and third we adopted halo mass estimates, obtained exploiting mock catalogs drawn from the Millennium Simulation \citep{springel05}.
In the latter case, we also included in our analysis a sample of galaxies in clusters from  WINGS+OmegaWINGS \citep{Fasano2006,Moretti2014,Gullieusik2015,Moretti2017} for which halo mass estimates were available, in order to extend the halo mass range covered. Such sample was assembled using similar cuts in magnitude and redshifts. 

Galaxies were classified based on the different features detected in their spectra (presence/strength of [OII], [OIII], H$\delta$ and H$\alpha$) into passive (PAS), post starburst (PSB) and emission line (EML) galaxies. We  studied the stellar population properties  of the different subpopulation with the aim to  shed light on the physical processes responsible for the star formation quenching. 

The main results can be summarized as follows. 
\begin{itemize}
\item  PSBs have properties (stellar mass, magnitude, color) intermediate between those of PAS and EML galaxies, suggesting that these objects are in a transitional stage between star-forming blue cloud galaxies and quiescent red sequence galaxies.  

\item
The environment drives the incidence of PSBs: while PSBs are always quite rare compared to the other populations, their  fraction decreases from clusters (4.5\%), to groups (3\%), to binaries (2\%), to single systems (1.1\%). 

\item
Trends are even more outstanding when using the estimates of halo masses: the PSB fraction increases monotonically with halo mass.

\item
{ There are hints that in binaries and singles the distribution of PSBs is skewed toward larger local density values than the other two populations. However, the K-S test is able to detect statistically meaningful differences only between PSBs and EMLs in groups and single systems.}
\end{itemize}

Our results can be explained advocating two different scenarios for the formation of PSB galaxies. 
In clusters and in (at least the massive) groups, the truncation of the star formation is most likely due to the interaction of the galaxy with the intracluster medium \citep[e.g.][]{Dressler1982,Couch1987,Poggianti2009,Paccagnella2017} or intergroup medium. Therefore, ram pressure stripping is the most likely driving mechanism. This mechanism is expected to be strongest in the most massive clusters \citep[e.g.][]{Gunn1972, Jaffe2015}, and this can explain the observed trend with halo mass. 

In binaries and singles the most obvious candidate mechanisms are close galaxy interactions or mergers, which can exhaust in a short time the available gas and thus give rise to a PSB spectrum. Our findings with the local density support this scenario. %As we did not detect any signs of disturbance in the morphological distribution of the PSBs little disruptive events, like minor mergers and accretion, are expected to be at play \citep[see also][]{Dressler2013}.

{Our findings imply that processes specific to the densest environments act on a larger fraction of galaxies than the processes producing PSB galaxies in the sparsest environments.}

\section*{Acknowledgements}
We thank the anonymous referee for their detailed comments that helped us to improve the paper. We are grateful to Joe Liske, Simon Driver and the whole MGC collaboration for making their dataset easily available, and to Rosa Calvi for her valuable work on the PM2GC. We also thanks Gabriella De Lucia for providing us with the mock catalogs from the Millennium Simulation. 
We acknowledge financial support from PRIN-SKA 2017 grant.
B.V. acknowledges the support from 
an  Australian Research Council Discovery Early Career Researcher Award (PD0028506).

%%%%%%%%%%%%%%%%%%%%%%%%%%%%%%%%%%%%%%%%%%%%%%%%%%



\bibliographystyle{apj}
\bibliography{biblio_paper3} % if your bibtex file is called example.bib

%%%%%%%%%%%%%%%%%%%%%%%%%%%%%%%%%%%%%%%%%%%%%%%%%%
%%%%%%%%%%%%%%%%%%%%%%%%%%%%%%%%%%%%%%%%%%%%%%%%%%

%%%%%%%%%%%%%%%%% APPENDICES %%%%%%%%%%%%%%%%%%%%%

\appendix

\section{The Halo mass catalogs} \label{sec:cat} 
We release a catalog with the halo mass estimates for the single, binary and group systems in the PM2GC, as computed in the previous section. Table \ref{tab:es_gr} provides the information for a subset of 10 groups, extracted from the main catalog; Table \ref{tab:es_bin} provides the information for a subset of 10 binary systems; Table \ref{tab:es_sin} provides the information for a subset of 10 single galaxies.   The full tables are available with the online version of the paper (see Supporting Information).

\begin{table}
\centering
\caption{Subsample of 10 galaxy groups with their properties. The full table is available with the online version of the paper (see Supporting Information). Columns: (1) PM2GC group serial number; (2, 3) geometric centre right ascension and declination in degrees; (4) Halo mass estimate; (5) Error on the halo mass estimate. \label{tab:es_gr}}
\begin{tabular}{ccccc}
\hline
 \multicolumn{1}{c}{ID\_GR} &
 \multicolumn{1}{c}{RA\_GR} &
 \multicolumn{1}{c}{DEC\_GR} &
 \multicolumn{1}{c}{$\log M_{halo}^{tot}$}  &
 \multicolumn{1}{c}{$\Delta(\log M_{halo}^{tot}$)}  \\
 \multicolumn{1}{c}{} &
 \multicolumn{1}{c}{(deg)} &
 \multicolumn{1}{c}{(deg)} &
 \multicolumn{1}{c}{($M_\odot$)}  &
 \multicolumn{1}{c}{($M_\odot$)}  \\
\hline
  1000  &170.3258  &-0.2097  &14.0    & 0.1   \\   
  1001  &197.1675  &0.0623   &12.53  &  0.03     \\  
  1002  &161.6724  &-0.0394  &12.95  &  0.05    \\   
  1003  &175.4628  &0.223    &12.80   &  0.04    \\   
  1004  &154.4308  &-0.0772  &13.53  &  0.09    \\   
  1006  &150.2518  &-0.1296  &12.85  &  0.05    \\   
  1007  &151.4264  &0.0807   &12.75  &  0.04    \\   
  1008  &155.8879 & -0.2185  &13.89  &  0.1   \\    
  1009  &169.9662 & 0.0633   &12.62  &  0.03    \\    
  1011  &150.259  & -0.1726 & 14.29  &  0.2   \\     
 \hline
\end{tabular}
\end{table}

\begin{table}
\centering
\caption{Subsample of 10 binary systems with their properties. The full table is available with the online version of the paper (see Supporting Information). Columns: (1) PM2GC binary serial number; (2, 3) PM2GC serial numbers of the two galaxies of the system; (4) Halo mass estimate; (5) Error on the halo mass estimate. \label{tab:es_bin}}
\begin{tabular}{ccccc}
\hline
 \multicolumn{1}{c}{ID\_BIN} &
 \multicolumn{1}{c}{IDMGC\_1} &
 \multicolumn{1}{c}{IDMGC\_2} &
 \multicolumn{1}{c}{$\log M_{halo}^{tot}$}  &
 \multicolumn{1}{c}{$\Delta(\log M_{halo}^{tot}$)}  \\
 \multicolumn{1}{c}{} &
 \multicolumn{1}{c}{(deg)} &
 \multicolumn{1}{c}{(deg)} &
 \multicolumn{1}{c}{($M_\odot$)}  &
 \multicolumn{1}{c}{($M_\odot$)}  \\
\hline
  1   &    19714  &   19621  &   11.98 &  0.01      \\
  2    &   29003   &  28996  &   11.62 &  0.01     \\
  3    &   9715    &  9645   &   12.39 &  0.02     \\
  4    &   23444   &  23429  &   12.24 &  0.01     \\
  5    &   11804   &  96374  &   11.66 &  0.01     \\
  6    &   61222   &  61196  &   11.54 &  0.01     \\
  7    &   27171   &  27176  &   12.15 &  0.01     \\
  8    &   64057   &  64105  &   12.79 &  0.04     \\
  9     &  16794   &  16813  &   11.44 &  0.01     \\
  10    &  64907   &  64887  &   11.34 &  0.01     \\
\\
 \hline
\end{tabular}
\end{table}

\begin{table}
\centering
\caption{Subsample of 10 single galaxies with their properties. The full table is available with the online version of the paper (see Supporting Information). Columns: (1) PM2GC galaxy serial number; (2) Halo mass estimate; (3) Error on the halo mass estimate. \label{tab:es_sin}}
\begin{tabular}{ccccc}
\hline
 \multicolumn{1}{c}{ID} &
 \multicolumn{1}{c}{$\log M_{halo}^{tot}$}  &
 \multicolumn{1}{c}{$\Delta(\log M_{halo}^{tot}$)}  \\
 \multicolumn{1}{c}{} &
 \multicolumn{1}{c}{($M_\odot$)}  &
 \multicolumn{1}{c}{($M_\odot$)}  \\
\hline
  61514 &  11.96 &  0.01     \\ 
  27280 &  11.63 &  0.01     \\
  11383 &  11.56 &  0.01     \\
  66778 &  11.41 &  0.01    \\ 
  52268 &  11.86 &  0.01     \\
  56234 &  11.58  &  0.01     \\
  16076 &  11.95 &  0.01     \\
  13606 &  11.44 &  0.01     \\
  57725 &  11.34 &  0.01     \\
  25661 &  11.63 &  0.01     \\
 \hline
\end{tabular}
\end{table}

%%%%%%%%%%%%%%%%%%%%%%%%%%%%%%%%%%%%%%%%%%%%%%%%%%
%%%%%%%%%%%%%%%%%%%%%%%%%%%%%%%%%%%%%%%%%%%%%%%%%%

% Don't change these lines
\bsp	% typesetting comment
\label{lastpage}
\end{document}